\documentclass[
twocolumn
,superscriptaddress,prd, twocolumn, nofootinbib, floatfix
]{revtex4-1}

\DeclareMathAlphabet{\mathcal}{OMS}{cmsy}{m}{n}
\usepackage{mathptmx}
\usepackage{tikz-cd}
\usepackage{color}
\usepackage{graphicx}
\usepackage{amsmath}
\usepackage{enumitem}
\usepackage{amssymb}
\usepackage{array}
\newcolumntype{L}[1]{>{\raggedright\let\newline\\\arraybackslash\hspace{0pt}}m{#1}}
\newcolumntype{C}[1]{>{\centering\let\newline\\\arraybackslash\hspace{0pt}}m{#1}}
\newcolumntype{R}[1]{>{\raggedleft\let\newline\\\arraybackslash\hspace{0pt}}m{#1}}
\usepackage{mathrsfs}
\usepackage{dsfont}
\usepackage{appendix}
\usepackage{multirow}
\usepackage{dsfont}
\usepackage{tikz}
\usepackage{amsfonts}
\usepackage{amssymb}
\usepackage{pgflibraryarrows}
\usepackage{pgflibrarysnakes}
\usetikzlibrary{decorations.pathmorphing}
\usepgflibrary{arrows.meta}

\DeclareMathOperator{\arcsinh}{arcsinh}

\usepackage{subfigure}
\usepackage{bbold}
\newcommand{\D}{\mathrm{d}}

\newcommand{\dd}{\dagger}

\newcommand{\infint}{\int_{-\infty}^\infty}
\usepackage{accents}
\usepackage{braket}

\usepackage{xtab,afterpage}
\usepackage{hyperref}
\usepackage{setspace,calc}
 \newcommand{\mz}{\color{green}}

\graphicspath{{Figures/}}
\allowdisplaybreaks

\definecolor{blueblue}{RGB}{21,47,181}

\hypersetup{linktoc=all,
    colorlinks, linkcolor={blueblue},
    citecolor={blueblue}, urlcolor={blueblue}
}

\begin{document}

\title{Schr\"odinger's cat for de Sitter spacetime}

\author{Joshua Foo}
\email{joshua.foo@uqconnect.edu.au}
\affiliation{Centre for Quantum Computation and Communication Technology, School of Mathematics and Physics, The University of Queensland, St. Lucia, Queensland, 4072, Australia}
\author{Robert B. Mann}
\affiliation{Department of Physics and Astronomy, University of Waterloo, Waterloo, Ontario, Canada, N2L 3G1}
\affiliation{Perimeter Institute, 31 Caroline St., Waterloo, Ontario, N2L 2Y5, Canada}
\author{Magdalena Zych}
\email{m.zych@uq.edu.au}
\affiliation{Centre for Engineered Quantum Systems, School of Mathematics and Physics, The University of Queensland, St. Lucia, Queensland, 4072, Australia}

\date{\today}

\begin{abstract}
{Quantum gravity is expected to contain descriptions of semiclassical spacetime geometries in quantum superpositions. To date, no framework for modelling such superpositions has been devised. Here, we provide a new phenomenological description for the response of quantum probes (i.e.\ Unruh-deWitt detectors) on a spacetime manifold in quantum superposition. By introducing an additional control degree of freedom, one can assign a Hilbert space to the spacetime, allowing it to exist in a superposition of spatial or curvature states. Applying this approach to static de Sitter space, we discover scenarios in which the effects produced by the quantum spacetime are operationally indistinguishable from those induced by superpositions of Rindler trajectories in Minkowski spacetime. The distinguishability of such quantum spacetimes from superpositions of trajectories in flat space reduces to the equivalence or non-equivalence of the field correlations between the superposed amplitudes.}
\end{abstract}
\maketitle

\section{Introduction}
Despite numerous attempts over the last century, a unified theory of quantum gravity remains elusive. Since the spacetime metric in general relativity is dynamical, standard approaches to quantisation cannot be applied to the gravitational field. Attempts at unification such as string theory \cite{gubser1998gauge,seiberg1999string,polchinski1998string,witten1995string} and loop quantum gravity \cite{rovelli2008loop,rovelli2014covariant,thiemann2003lectures} have addressed the problem from a top-down perspective. However, the well-known difficulties within these approaches have led to suggestions that spacetime could be fundamentally classical, or that the quantisation of gravity is an ill-posed question to begin with \cite{dyson2014graviton}. For pertinent discussions on the necessity of quantising the gravitational field, see \cite{garay1995quantum,carlip2008quantum} and references therein.

Any expected theory combining spacetime with the indeterminacy of quantum theory should admit such notions as a superposition of semiclassical geometries. Christodoulou and Rovelli \cite{christodoulou2019possibility} have argued that a recent experimental proposal, involving the gravitationally-induced entangling of two mesoscopic particles (each in a spatial superposition state) \cite{bose2017spin,marletto2017gravitationally}, would provide evidence for a system described by a superposition of spacetime geometries. So far the focus has been on effects produced by spatial superpositions of massive objects. However, the resulting metrics only differ by a global coordinate transformation and are thus diffeomorphic. Such scenarios can equivalently be described as arising in a single classical spacetime where quantum systems are prepared and measured in expropriate quantum states. Quantum systems residing in background metrics defined by parameters in superposition, or even superpositions of different classes of background metrics, has not been studied. 

In this paper, we propose a new phenomenological description of a spacetime metric in a genuine quantum superposition. Rather than offering a fully quantum-gravitational origin for this phenomenon, we work within the paradigm of quantum field theory in curved spacetime (QFT-CS), using a theoretical tool known as the Unruh-deWitt (UdW) detector \cite{unruh1984happens} to probe such superpositions.  UdW detectors have been utilized widely in relativistic and gravitational settings, for example in the study of quantum fields and their entanglement structure in black hole spacetimes \cite{ng2014unruh,ng2017over,ng2018new,henderson2019btz,hodgkinson2012static} and higher-dimensional topologies \cite{hodgkinson2012often}. We adopt a recent framework \cite{foo2020unruhdewitt,barbado2020unruh} which extends the model to include quantum-controlled superpositions of semiclassical detector trajectories. As suggested in \cite{foo2020unruhdewitt}, this would allow (for example) the detector to reside in a superposition state of different radial distances from a black hole horizon, in which it would be expected to perceive a superposition of redshifts induced by the local curvature, and respond to Hawking radiation as if it were in a coherent {\mz `}superposition{\mz '} of thermal states. 

In the first application of this novel idea, we study the experience of UdW detectors residing on a background de Sitter spacetime in superposition. We beging by modelling the detector as traveling in a superposition of spatially translated worldlines in the static patch of de Sitter spacetime. We demonstrate that this scenario is diffeomorphic to that in which the detector resides on a manifold in a quantum superposition of spatial translations \cite{zych2018relativity}. Our main result is the application of the model to phenomenologically describe de Sitter spacetime in a superposition of curvatures. The detector interacts with quantum fields defined on a background spacetime described by a superposition of states, each associated with a different value (i.e.\ superposed) of the de Sitter curvature. This novel description bears an analogy to the superposition of `semiclassical' coherent states in quantum information. Unlike superpositions of spatial degrees of freedom, there is no diffeomorphism that maps the metric in superposition to a single metric with a classical value of curvature.  

In each scenario, we calculate the response of the detector as it interacts with a conformally coupled massless scalar field in static de Sitter spacetime. To leading order in perturbation theory, we discover particular scenarios where the detector response corresponds with that produced by superpositions of uniformly accelerated trajectories in Minkowski spacetime. Due to the well-known conformal relationship between the Rindler and de Sitter geometries, it is on one hand unsurprising that such a correspondence exists. On the other hand, this raises the interesting question of how one can operationally distinguish between genuine superpositions of spacetime metrics from mere superpositions of trajectory states of an UdW detector.

We organise this paper as follows: in Sec.\ \ref{sec:II}, we review the UdW detector model with the inclusion of a control degree of freedom, which allows the detector to travel in quantum superpositions of trajectories, or equally initialises the background spacetime in an arbitrary superposition of quantum states. In Sec.\ \ref{sec:III}, we demonstrate the diffeomorphic invariance of a superposition of detector trajectory states on a classical metric compared with a classical detector trajectory on a background in a superposition of spatial states. We apply the model to static de Sitter spacetime in Sec.\ \ref{sec:V} and \ref{sec:VI}, studying operationally the effect of metrics in superpositions of spatial and curvature states on the behaviour of quantum detectors. We discuss the question of how quantum probes may differentiate genuine superpositions of spacetime metrics from flat spacetime trajectories in Sec.\ \ref{sec:VII}, before offering some conclusions in Sec.\ \ref{sec:VIII}. Throughout, we utilise natural units, $\hslash = k_B = c = G = 1$. 

\section{UdW detectors in superposition}\label{sec:II}

In the original formulation of the UdW detector model with quantum control, the detector is modelled as interacting with the quantum field on a classical spacetime background in a superposition of trajectory states. Such a system can be described in the tensor product Hilbert space, $\mathcal{H} = \mathcal{H}_T \otimes \mathcal{H}_\text{UdW} \otimes \mathcal{H}_F$ where $\mathcal{H}_T$, $\mathcal{H}_\text{UdW}$ and $\mathcal{H}_F$ are associated with the trajectory, detector, and field degrees of freedom respectively.  We are interested here in describing the spacetime itself with quantised degrees of freedom.
Thus, the Hilbert space can be decomposed as $\mathcal{H} = \mathcal{H}_S \otimes \mathcal{H}_\text{UdW} \otimes \mathcal{H}_F$, where $\mathcal{H}_S$, $\mathcal{H}_\text{UdW}$ and $\mathcal{H}_F$ are now associated with the spacetime, detector and field degrees of freedom. As we demonstrate in Sec.\ \ref{sec:III}, the two representations are essentially equivalent for spatial superpositions. Finally, we note that in our analysis in Sec.\ \ref{seckms}, both the spacetime and the trajectories are assigned quantum states.

The standard UdW model considers a point-like two-level system whose internal states $|g\rangle, |e\rangle$ couple to the massless, scalar field $\hat{\Phi}(\mathsf{x}(\tau))$. We assume that the detector is initially in its ground state, $|g\rangle$, and that the field is in the conformally coupled vacuum state $|0\rangle_F$ of de Sitter spacetime \cite{birrell1984quantum}, that is
\begin{align}
    |\Psi\rangle_{FD} &= |g\rangle \otimes |0\rangle_F .
\end{align}
The conformal vacuum is a natural choice, since it is a coordinate-independent vacuum state bearing a close analogy to the Minkowski vacuum in flat spacetime \cite{birrell1984quantum}. We introduce a control degree of freedom, $|\chi\rangle$, to the detector-field-spacetime system, leaving the initial state to be 
\begin{align}
    |\Psi\rangle_{CFD} &= |\chi \rangle \otimes |g \rangle \otimes | 0\rangle_F
\end{align}
where
\begin{align}
    |\chi \rangle &= \frac{1}{\sqrt{N}} \sum_{i=1}^N |i\rangle_C
\end{align}
and $|i\rangle_C$ are orthogonal  states. As alluded to, one may interpret the control as being attached to the detector itself, in which case it governs the trajectories which the detector traverses in superposition. However we are motivated by the possibility of describing the detector travelling on a single trajectory with the spacetime in a superposition state. Without postulating the (quantum-gravitational) mechanism by which one prepares a superposition of a gravitational source of curvature, we show in this article that one can equally take $|\chi\rangle$ as controlling the quantum state of the background spacetime. In such a scenario, the amplitudes of the superposition could correspond to the orthogonal `curvature states' of an expanding spacetime, or the `quantised mass states' of a black hole \cite{arrasmith2019,wood2020minimum}. 

Returning now to the UdW detector model with a quantum control, we introduce the interaction Hamiltonian,
\begin{align}
    \hat{H}_\text{int.} (\tau) &= \sum_{i=1}^N \hat{\mathcal{H}}_i(\tau) \otimes |i\rangle\langle i |_C
\end{align}
where
\begin{align}\label{eq:5}
    \hat{\mathcal{H}}_i(\tau) &= \lambda\hat{\sigma}(\tau) \eta_i(\tau ) \hat{\Phi}\big( \mathsf{x}_i(\tau ) \big) 
\end{align}
governs the interaction along the worldline $\mathsf{x}_i(\tau)$ of the superposition. In Eq.\ (\ref{eq:5}), $\lambda$ is a weak coupling constant, $\eta_i(\tau)$ are switching functions, $\hat{\sigma}(\tau) = |e\rangle\langle g| e^{i\Omega \tau} + \text{H.c}$ is the interaction picture ladder operator between the states $|g\rangle, |e\rangle$ with energy gap $\Omega$. The evolution of the system in the interaction picture is given by \begin{align}
    \hat{U} |\Psi\rangle_{CFD} = \mathcal{T} \exp \bigg\{ -i\int\D t \: \Big( \frac{\D \tau}{\D t}H_\text{int.}(\tau) \Big) \bigg\} |\Psi\rangle_{CFD}
\end{align}
where $\mathcal{T}$ denotes time-ordering. The time-evolution operator takes the form, 
\begin{align}\label{eq6}
    \hat{U} &= \sum_{i=1}^N \hat{U}_i \otimes |i\rangle\langle i|_C 
\end{align}
where, to leading order in perturbation theory
\begin{align}
    \hat{U}_i &= 1 - i \lambda\int\D \tau \:\mathcal{\hat{H}}_i(\tau) + \mathcal{O}(\lambda^2).
\end{align}
Using Eq.\ (\ref{eq6}), we can time-evolve the initial state to obtain
\begin{align}
    \hat{U} |\Psi\rangle_{CFD} &= \frac{1}{\sqrt{N}} \sum_{i=1}^N \hat{U}_i  |i\rangle_C |0\rangle_F|g\rangle .
\end{align}
Following \cite{foo2020unruhdewitt,foo2020thermality}, we are interested in the conditional state of the detector given the control is measured in some fixed state, which we take to be its initial state, $|\chi\rangle$. This yields
\begin{align}\label{10}
    \langle \chi | \hat{U} |\Psi\rangle_{CFD} \equiv |\Psi'\rangle_{FD} &= \frac{1}{N} \sum_{i=1}^N \hat{U}_i | 0\rangle_F | g\rangle .
\end{align}
From this, one can obtain the final state of the detector by tracing out the field degrees of freedom, which to leading order in the coupling constant $\lambda$, is given by the density matrix
\begin{align}
    \hat{\rho}_D &= \begin{pmatrix} 1 - \mathcal{P}_D & 0 \\ 0 & \mathcal{P}_D \end{pmatrix} + \mathcal{O}(\lambda^4) 
\end{align}
where
\begin{align}\label{probability}
    \mathcal{P}_D &= \frac{\lambda^2}{N^2} \sum_{i,j=1}^N \mathcal{P}_{ij,D} = \frac{\lambda^2}{N^2} \left\{ \sum_{i=j}^N \mathcal{P}_{ij,D} + \sum_{i\neq j} \mathcal{P}_{ij,D} \right\} 
\end{align}
is the transition probability of the detector (or fraction of excited detectors within an identically prepared ensemble) conditioned on the measurement of the control system in the asymptotic future. The individual contributions take the form
\begin{align}\label{probs2}
    \mathcal{P}_{ij,D} &= \int\D \tau \int\D\tau' \chi_i(\tau) \overline{\chi}_j(\tau') \mathcal{W}^{ji}\big( \mathsf{x}_i, \mathsf{x}_j' \big)
\end{align}
where we have defined $\chi(\tau) = \eta(\tau) e^{-i\Omega\tau}$, while
\begin{align}\label{Wightmanequation} 
\mathcal{W}^{ji}\big( \mathsf{x}_i, \mathsf{x}_j'\big) := \langle 0 | \hat{\Phi}(\mathsf{x}_i ) \hat{\Phi}(\mathsf{x}_j' ) |0\rangle_F
\end{align}
are field correlation (Wightman) functions pulled back to the trajectories $\mathsf{x}_i(\tau), \mathsf{x}_j'(\tau')$ \cite{birrell1984quantum}, associated with the $i$-$j$th amplitude of the superposition. The $i$th trajectory is associated with the $|i\rangle_C$ state of the control system. Unlike a detector travelling along a single, classical trajectory, $\mathcal{P}_D$ now features `nonlocal' correlation functions between each respective pair of amplitudes in the superposition, $i \neq j$. 

For the switching function, we consider a Gaussian, $\eta (\tau) = \exp(-\tau^2/2\sigma^2)
$, where $\sigma$ is a characteristic timescale of the interaction. For stationary trajectories, the Wightman functions only depend on $s = \tau - \tau'$ and the outer integral of Eq.\ (\ref{probs2}) can be evaluated exactly, yielding the simplified expression,
\begin{align}
   \mathcal{P}_D &= \frac{\lambda^2\sqrt{\pi}\sigma}{N^2} \sum_{i,j=1}^N \int\D s \:e^{-s^2/4\sigma^2}e^{-i\Omega s}\mathcal{W}^{ji}(s) .
\end{align}
In these time-independent scenarios, it will be useful to work with the normalised transition probability
\begin{align}\label{response}
    \mathcal{F}(\Omega) &= \frac{\mathcal{P}_D}{\lambda^2\sqrt{\pi}\sigma},
\end{align}
which we refer to as the response function. In the infinite-interaction time limit, $\sigma\to \infty$, the response function becomes
\begin{align}\label{14}
    \mathcal{F}(\Omega) &= \frac{\lambda^2}{N^2} \sum_{i,j=1}^N \infint\D s \:e^{-i\Omega s }\mathcal{W}^{ji} (s).
\end{align}
For non-stationary trajectories where analytic solutions are intractable, we can calculate the instantaneous transition rate of the detector, measured at the proper time $\tau$ while the interaction is still on. Following \cite{foo2020unruhdewitt,foo2020thermality}, and taking the infinite interaction-time limit $(\sigma\to \infty$), one obtains for the instantaneous transition rate, 
\begin{align}\label{past}
    \dot{\mathcal{P}}_D &=  \frac{2\lambda^2}{N^2} \text{Re}\sum_{i,j=1}^N \int_0^\infty \D s \:e^{-i\Omega s}\mathcal{W}^{ji}(\tau,\tau-s)
\end{align}
where $\dot{\mathcal{P}}_D = \D \mathcal{P}_D/\D \tau$.

\section{Diffeomorphic invariance of superpositions of spatial translations}\label{sec:III}
Equations (\ref{14}) and (\ref{past}) characterise the detector's response as it travels along a given trajectory or superposition of trajectories in spacetime. These are essentially Fourier transforms of the Wightman functions, Eq.\ (\ref{Wightmanequation}), which encode the field correlations between $\hat{\Phi}\big(\mathsf{x}_i(\tau) \big)$ and $\hat{\Phi}\big(\mathsf{x}_j(\tau) \big)$. In the language of superposed UdW detectors traveling on a classical background spacetime, it is natural to view $\mathsf{x}_i(\tau)$ and $\mathsf{x}_j(\tau)$ as the superposed trajectories of the detector itself. Alternatively, such a scenario can be mapped via a global coordinate transformation so that the detector travels along a single classical trajectory, with the spacetime in a superposition of spatial translations \cite{zych2018relativity}. 

To illustrate this, let us consider a detector in a superposition of two trajectories. One can associate the basis states $|i\rangle_C$ in the control superposition with the physical trajectories which the detector traverses. For example, we take
\begin{align}
    |1 \rangle_C &\Rightarrow |\xi \rangle \vphantom{\frac{1}{2}} \\
    |2 \rangle_C &\Rightarrow |\xi + \mathcal{L} \rangle 
\end{align}
which we refer to as the trajectory states, where $\xi\equiv \xi(\tau)$ denotes the worldline of the first trajectory and $\mathcal{L} \equiv \mathcal{L}(\tau)$ is some time-dependent function that relates the coordinates of the trajectory basis states. For fixed $\mathcal{L}$, this merely describes a constant spatial translation, which we assume for simplicity here.  

A coordinate transformation  
relating the trajectory basis states can be
enacted via a unitary transformation $\hat{\mathcal{T}}(\mathcal{L})$
\begin{align}\label{basis}
    |\xi + \mathcal{L} \rangle \otimes |0\rangle_F &= \hat{\mathcal{T}}(\mathcal{L}) |\xi\rangle \otimes |0\rangle_F \vphantom{\frac{1}{2}}\\
    &= \hat{\mathcal{T}}_\xi(\mathcal{L}) | \xi \rangle \otimes \hat{\mathcal{T}}_\phi(\mathcal{L}) | 0\rangle_F 
\end{align}
where $\hat{\mathcal{T}}(\mathcal{L}) := \hat{\mathcal{T}}_\phi(\mathcal{L}) \otimes \hat{\mathcal{T}}_\xi(\mathcal{L})$, i.e.\ the unitary $\hat{\mathcal{T}}(\mathcal{L})$
acts on the trajectory state (generating spatial translations through $\hat{\mathcal{T}}_\xi(\mathcal{L})$) and also on the field degrees of freedom, since they possess a dependence on the spacetime coordinates. When calculating transition amplitudes, $\hat{\mathcal{T}}_\phi(\mathcal{L})$ generates unitary coordinate transformations at the level of the field operator, 
\begin{align}
    \hat{\Phi}(\xi + \mathcal{L}) &:= \hat{\mathcal{T}}_\phi(\mathcal{L})^\dd \hat{\Phi}(\xi)  \hat{\mathcal{T}}_\phi(\mathcal{L}).
\end{align}
Due to the symmetries of general relativity, any arbitrary coordinate transformation of the spacetime manifold has an associated unitary representation that transforms quantum states defined on that manifold in a corresponding way. Our goal here is to demonstrate that the dynamical evolution of the detector-field-spacetime system (and hence observables measured by the detector) is diffeomorphic invariant under these unitary transformations of the field operators. 

Using Eq.\ (\ref{basis}), the initial state of the system  can be expressed as
\begin{align}\label{21}
    |\Psi\rangle_{TFD} &= \frac{1}{\sqrt{2}} \big( |\xi \rangle + | \xi + \mathcal{L} \rangle \big) |0\rangle_F | g \rangle \\
    &= \frac{1}{\sqrt{2}} \big(\mathds{1} + \hat{\mathcal{T}}(\mathcal{L}) \big) |\xi \rangle | 0\rangle_F  |g\rangle .
\end{align}
Thus, the time-evolution of the state is given by 
\begin{align}\label{22}
    \hat{U} |\Psi\rangle_{TFD} &= \frac{1}{\sqrt{2}} \big( \hat{U} + \hat{U}\hat{\mathcal{T}}(\mathcal{L}) \big) |\xi\rangle |0\rangle_F |g\rangle .
\end{align}
From Eq.\ (\ref{22}), we see that the detector motion has been mapped to a single trajectory, $\xi$, and the detector-field dynamics are now enacted via a superposition of unitaries. It is now the field operators contained in $\hat{U}$ which carry the superposition dynamics of the system.

Measuring the trajectory state in the modified basis $|\chi\rangle = ( \mathds{1} + \hat{\mathcal{T}}(\mathcal{L}) ) |\xi \rangle/\sqrt{2}$ yields the conditional detector-field state
\begin{align}
    |\Psi\rangle_{FD} &= \frac{1}{2} \Big( \langle \xi |\hat{U} |\xi\rangle  + \langle \xi |\hat{U} \hat{\mathcal{T}}(\mathcal{L}) |\xi\rangle  + \langle \xi |\hat{\mathcal{T}}(\mathcal{L})^\dd \hat{U} |\xi\rangle \nonumber \\
    & + \langle \xi|\hat{\mathcal{T}}(\mathcal{L})^\dd \hat{U} \hat{\mathcal{T}}(\mathcal{L})|\xi\rangle \Big) \otimes  |0\rangle_F \otimes |g\rangle \vphantom{\frac{1}{2}} .
\end{align}
Previously, we have considered the time-evolution operator as having components acting on all of the trajectory states of the detector, 
\begin{align}
    \hat{U} \equiv \sum_\text{$\zeta =$  \text{paths}} \hat{U} ( \zeta ) \otimes | \zeta \rangle \langle \zeta | .
\end{align}
However now we only have the trajectory $|\xi\rangle$, so 
\begin{align}\label{26}
    \hat{U} &= \hat{U} (\xi ) \otimes | \xi \rangle \langle \xi|. 
\end{align}
where $\hat{U}(\xi)$ acts on the detector-field Hilbert space. Using Eq.\ (\ref{26}), it is straightforward to show that $\langle \xi | \hat{U} \hat{\mathcal{T}}(\mathcal{L} ) |\xi \rangle = \langle \xi | \hat{\mathcal{T}}(\mathcal{L})^\dd \hat{U} |\xi \rangle = 0$ while 
\begin{align}
    \langle \xi | \hat{U} | \xi \rangle &= \hat{U} (\xi ) \\ \label{28}
    \langle \xi | \hat{\mathcal{T}}(\mathcal{L})^\dd \hat{U} \hat{\mathcal{T}}( \mathcal{L} ) | \xi \rangle &= \hat{\mathcal{T}}_\phi (\mathcal{L})^\dd \hat{U}(\xi) \hat{\mathcal{T}}_\phi(\mathcal{L}),
\end{align}
where Eq.\ (\ref{28}) is simply equal to $\hat{U} ( \xi + \mathcal{L})$. This leaves the detector-field system in the state
\begin{align}\label{29}
    |\Psi\rangle_{FD} &= \frac{1}{2} \big( \hat{U}(\xi) + \hat{U}(\xi + \mathcal{L}) \big) |0\rangle_F |g \rangle .
\end{align}
We see clearly that the detector, now traversing the single classical trajectory $\xi \equiv \xi(\tau)$, undergoes dynamics in which the field operators are enacted in a superposition state of spatial translations (i.e.\ evaluated along two different trajectories). Of course, Eq.\ (\ref{29}) is exactly Eq.\ (\ref{10}), viewed from a different quantum reference frame \cite{giacomini2019}. 

For completeness, we can write explicitly the expressions for the time-evolution operators, namely
\begin{align}
    \hat{U}(\xi) &= 1 - i \int\D \tau \: \hat{\mathcal{H}}_i( \xi )  + \mathcal{O}(\lambda^2 ) \\
    \hat{U}(\xi + \mathcal{L} ) &= 1 - i\int\D \tau \:\hat{\mathcal{H}}_i(\xi + \mathcal{L}) + \mathcal{O}(\lambda^2).
\end{align}
Applying these to the detector-field state in Eq.\ (\ref{29}) yields, after tracing out the field states,  the conditional transition probability 
\begin{align}\label{diffeopd}
    \mathcal{P}_D &= \frac{\lambda^2}{4} \left\{ \sum_{i=j} \mathcal{P}_{ij,D} + \sum_{i\neq j} \mathcal{P}_{ij,D}  \right\}
\end{align}
 where
\begin{align}\label{diffeopd-2}
    \mathcal{P}_{ij,D} &= \int\D \tau \int\D \tau' \chi(\tau) \overline{\chi}(\tau') \langle 0 | \hat{\Phi}(i) \hat{\Phi}(j') | 0\rangle_F 
\end{align}
are the individual contributions evaluated between the trajectories $i = \xi, \xi + \mathcal{L}$ and $j' = \xi', \xi' + \mathcal{L}$. 
\begin{figure}[h]
    \centering
    \includegraphics[width=1\linewidth]{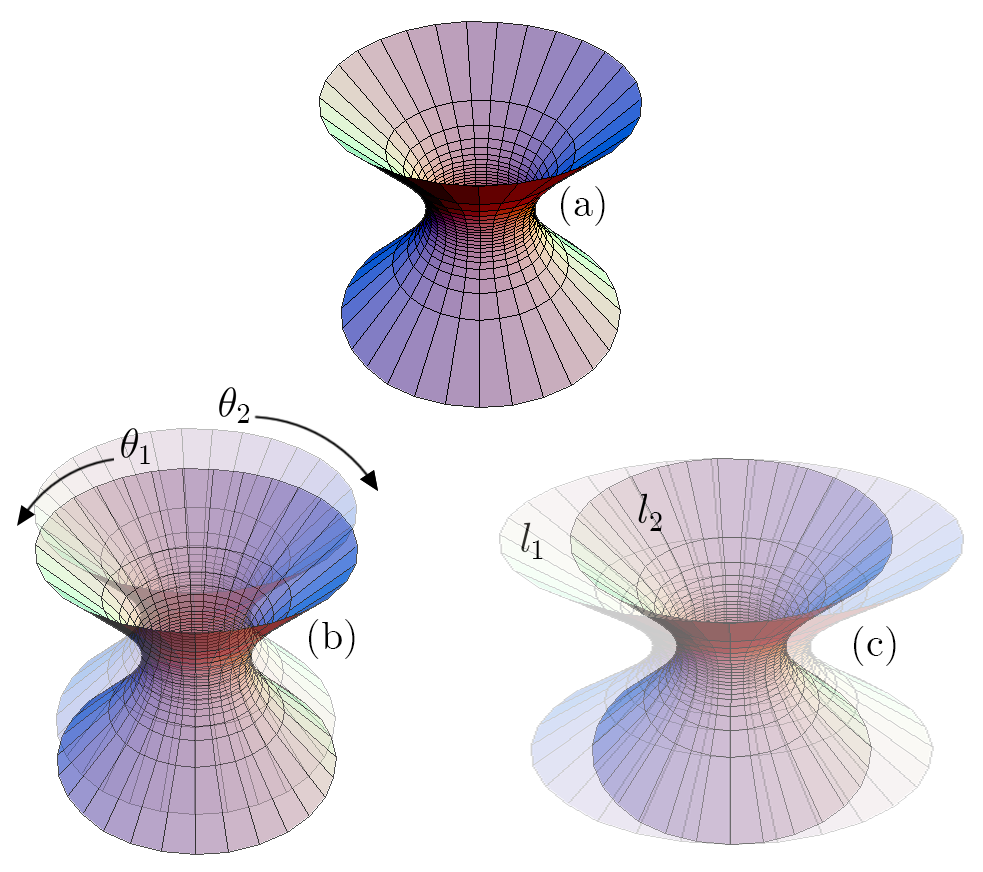}
    \caption{Visualisation of (a) classical de Sitter spacetime, (b) the de Sitter hyperboloid in a superposition of angular separations, and (c) de Sitter spacetime in a superposition of curvatures. Naturally, two dimensions have been suppressed. In (b) we have offset the two hyperboloids for ease of visualisation; of course, the angular translation is within the 5-dimensional embedding space.}
    \label{fig:visualisation}
\end{figure}
Comparing Eq.\ (\ref{diffeopd}) with Eq.\ (\ref{probability}), it is clearly seen that the transition probability is equivalent between the two perspectives. 
More generally, for a detector prepared in a superposition of trajectory states on a fixed background spacetime (i.e.\ in a quantum reference frame), there always exists a complementary scenario in which the detector traverses a single worldline (i.e.\ a classical reference frame) and the spacetime is in a quantum superposition of spatial states. Such a scenario is visualised in Fig.\ \ref{fig:visualisation}(b) for a superposition of angular rotations in the 5-dimensional embedding space. Observables like the transition probability of the detector are diffeomorphic invariant when transforming between these perspectives. As we discuss in Sec.\ \ref{sec:VI}, this invariance does not apply to situations involving superpositions of curvature, since two unique solutions to Einstein's equations in superposition cannot be mapped to a single classical background. This scenario, visualised in Fig.\ \ref{fig:visualisation}(c), unambiguously represents a genuine quantum superposition of spacetime metrics that has no dual representation in terms of a superposition of detector trajectories.

\section{Static de Sitter and Rindler geometries}\label{sec:IV}
Before studying the specific detector-de Sitter spacetime superpositions alluded to previously, we review the necessary geometric and field-theoretic details for our calculations. In this paper we study superpositions of static de Sitter spacetime \cite{de1917over,de1917over2}, which is particularly convenient for attaining simple analytic results due to its constant curvature. Throughout, we also make comparisons to analogous scenarios for accelerated trajectories in flat spacetime, and so we also briefly review the geometric details of Rindler space below. 

\subsection{Static de Sitter spacetime}
de Sitter spacetime is important to quantum field theorists due to its high degree of symmetry (along with anti-de Sitter space, it is the only maximally symmetric and curved solution to Einstein's equations) and its application to cosmology, where the inclusion of a positive cosmological constant $\Lambda$ makes it a simple description of an expanding spacetime \cite{griffiths2009exact,barrabes2008einstein}. The de Sitter manifold can be visualised as the hyperboloid \cite{de1917over,de1917over2},
\begin{align}
    -Z_0^2 + Z_1^2 + Z_2^2 + Z_3^2 + Z_4^2 &= \frac{1}{l^2}
\end{align}
where $l = \sqrt{\Lambda/3}$ is known as the de Sitter length and $\Lambda$ is the cosmological constant (here, we follow the convention of \cite{salton2015acceleration,ver2009entangling,kukita2017entanglement,nambu2013entanglement} and associate $l$ with the expansion rate of the spacetime, rather than its inverse as utilised elsewhere). Since the expansion rate is related to the Ricci scalar by $R = 12l^2$ \cite{akhtar2019open}, we will henceforth refer to the expansion rate and spacetime curvature interchangeably. The constancy of $l$ makes it a simple case to study when considering superpositions of curvature states. 

The de Sitter hyperboloid is embedded within a flat, five-dimensional Minkowski spacetime, 
\begin{align}\label{Zmetric}
    \D s^2 &= - \D Z_0^2 + \D Z_1^2 + \D Z_2^2 + \D Z_3^2 + \D Z_4^2 .
\end{align}
The hyperboloid can be parametrised using different coordinate systems. In this paper we consider detectors confined to the static patch of de Sitter spacetime, which can be described using the spherically symmetric coordinates $(t,r,\theta,\phi)$, yielding the corresponding line element
\begin{align}\label{metric}
    \D s^2 &= - f(r) \D t^2 + \frac{\D r^2}{f(r)} +r^2\D \Omega_2^2
\end{align}
where
\begin{align}
    f(r) &= 1 - l^2r^2
\end{align}
and $\D\Omega_2^2 = \D\theta^2 + \sin^2\theta\D \phi^2$, and $t \in(-\infty, \infty)$, $r \in[0,1/l)$, $\theta \in [0,\pi]$, and $\phi \in [0,2\pi]$. The null hypersurface at $r = 1/l$ forms a cosmological horizon. Furthermore, a test particle at fixed $(R_D,\theta,\phi)$ has non-zero four-acceleration with magnitude \begin{align}
\alpha =  \frac{lR_D}{\sqrt{l^{-2} - R_D^2}}.
\end{align}
The acceleration of the particle effectively counteracts the expansion of the spacetime, and so we refer to such trajectories as \textit{static} trajectories. Clearly, as one approaches the cosmological horizon $r \to 1/l$, the acceleration becomes unbounded, while as $l\to 0$ or $r\to 0$, $\alpha$ vanishes. 

For a conformally coupled massless scalar field, the Wightman functions take the form \cite{birrell1984quantum,polarski1989hawking}
\begin{align}\label{39}
    \mathcal{W}_\text{dS}\big(\mathsf{x}(\tau),\mathsf{x}'(\tau') \big) &= - \frac{1}{4\pi^2} \frac{1}{\sigma(\mathsf{x},\mathsf{x}') - i\varepsilon }
\end{align}
where
\begin{align}
    \sigma(\mathsf{x},\mathsf{x}') &= (Z_0 - Z_0')^2 - (Z_1 - Z_1')^2 - (Z_2 - Z_2')^2 \nonumber \\
    & - (Z_3 - Z_3')^2 - (Z_4-Z_4')^2
\end{align}
is the geodesic distance between $\mathsf{x}$ and $\mathsf{x}'$ in the embedding space, and $\varepsilon$ is an infinitesimal regularisation constant. Note that the unprimed and primed coordinates signify that the coordinates are evaluated at the times $t$, $t'$ respectively and that the coordinate $t$ is related to the proper time of the detector via $t = \tau /\sqrt{f(R_D)}$. Using Eq.\ (\ref{39}) and the parametrisation
\begin{align}
    Z_0 &= \sqrt{l^{-2} - R_D^2} \sinh(l t ) \\ 
    Z_1 &= \sqrt{l^{-2} - R_D^2 } \cosh(lt ) \\
    Z_2 &= R_D \cos\theta \vphantom{\sqrt{l^{-2} - R_D^2}} \\
    Z_3 &= R_D \sin\theta\cos\phi \vphantom{\sqrt{l^{-2} - R_D^2}} \\
    Z_4 &= R_D \sin\theta\sin\phi \vphantom{\sqrt{l^{-2} - R_D^2}}
\end{align}
for a detector at constant $(R_D, \theta, \phi)$, the Wightman functions take the form 
\begin{align}\label{wightmandesitter}
    \mathcal{W}_\text{dS}(s) &= - \frac{\kappa^2}{16\pi^2} \frac{1}{\sinh^2 \big( \kappa s/2 - i\varepsilon\big)}
\end{align}
where $s = \tau - \tau'$ is the proper time difference and $\kappa  = (l^{-2} - R_D^2)^{-1/2}$. In the infinite interaction-time limit, the response function of an UdW detector on a static worldline can be shown to be \cite{gibbons1977cosmological}
\begin{align}\label{staticresponse}
    \mathcal{F}_\text{dS}^{\sigma\to\infty} (\Omega) &= \frac{\Omega}{2\pi} \frac{1}{e^{2\pi\Omega/\kappa} - 1}.
\end{align}
Equation (\ref{staticresponse}) is a thermal spectrum satisfying the detailed balance form of the Kubo-Martin-Schwinger (KMS) condition \cite{haag1967equilibrium}, 
\begin{align}
    \mathcal{R}(\Omega):= \frac{\mathcal{F}(\Omega)}{\mathcal{F}(-\Omega)} = e^{-2\pi\Omega/\kappa} 
\end{align}
with temperature $\kappa(2\pi)^{-1}$ and we have defined $\mathcal{R}(\Omega)$ as the excitation-to-deexcitation ratio of the UdW detector's internal states. This is to say that the detector perceives the conformal vacuum to be a thermal bath radiating at the Gibbons-Hawking temperature
\begin{align}
    T_\text{dS} = \frac{\kappa}{2\pi} &= \frac{1}{2\pi \sqrt{l^{-2} - R_D^2}} = \frac{\sqrt{l^2 + \alpha^2}}{2\pi }.
\end{align}
This describes the combined effects of both the cosmological expansion of the spacetime, producing Gibbons-Hawking radiation due to the presence of a horizon, as well as the production of Rindler particles induced by the intrinsic acceleration of the detector, due to the Unruh effect \cite{casadio2011unruh,rabochaya2016quantum}. 



\subsection{Rindler spacetime}
Throughout this paper, we make comparisons between superpositions of de Sitter spacetime and uniformly accelerated trajectory states in flat Minkowski space. Such worldlines are characterised by Rindler coordinates, which we briefly review here. 

Beginning with flat Minkowski spacetime, Eq.\ (\ref{Zmetric}), one can make the following boost
\begin{align}\label{boost1}
    Z_0 &\mapsto Z_0 \cosh \beta + Z_1 \sinh \beta \\ \label{boost2}
    Z_1 &\mapsto Z_0 \sinh \beta + Z_1 \cosh \beta
\end{align}
where $\beta$ is the boost parameter, which leaves the spacetime invariant \cite{Crispino:2007eb}. This boost invariance motivates the following coordinate transformation,
\begin{align}
    Z_0 &= \xi \sinh \tau \\
    Z_1 &= \xi \cosh \tau 
\end{align}
for which the metric takes the form
\begin{align}
    \D s^2 &= - \xi^2 \D \tau^2 + \D \xi^2 + \D Z_2^2 + \D Z_3^3 + \D Z_4^2 .
\end{align}
Worldlines with constant $\xi, Z_i$ describe the trajectories produced by the boost transformation, Eq.\ (\ref{boost1}) and (\ref{boost2}), and possess the uniform proper acceleration $\xi^{-1}$. The further coordinate transformation $\xi = \kappa_M^{-1} \exp ( \kappa_M \bar{\xi} )$, $\tau = \kappa_M \bar{\tau}$, or 
\begin{align}
    Z_0 &= \kappa_M^{-1} e^{\kappa_M \bar{\xi}} \sinh (\kappa_M \bar{\tau}) \\
    Z_1 &= \kappa_M^{-1} e^{\kappa_M \bar{\xi}} \cosh (\kappa_M \bar{\tau}) 
\end{align}
yields the metric \cite{griffiths2009exact}
\begin{align}
    \D s^2 &= e^{2\kappa_M \bar{\xi}} \big( \D \bar{\tau}^2 - \D \bar{\xi}^2 \big) - \D Z_2^2 - \D Z_3^2 - \D Z_4^2.  
\end{align}
This coordinate system is especially useful since the worldline with $\bar{\xi} = 0$,
\begin{align}\label{rindlercoords1}
    Z_0 &= \kappa_M^{-1} \sinh( \kappa_M \bar{\tau} ) \\ \label{rindlercoords2}
    Z_1 &= \kappa_M^{-1} \cosh (\kappa_M \bar{\tau})  \\
    Z_i &= 0 \vphantom{\kappa_M^{-1}}
\end{align}
has constant proper acceleration $\kappa_M$. These trajectories produce hyperbolae in the region $|Z_1|>|Z_0|$ in the Cartesian form of the Minkowski coordinates, Eq.\ (\ref{Zmetric}), known as the right Rindler wedge \cite{Crispino:2007eb}. 

In flat spacetime, the Wightman functions can be obtained via the usual mode-sum expansion of the fields, yielding the result \cite{birrell1984quantum}
\begin{align}\label{Wightmanflat2}
    \mathcal{W}_M \big(\mathsf{x}(\tau), \mathsf{x}'(\tau' ) \big)  &= \frac{-1/4\pi^2}{(Z_0 - Z_0' - i\varepsilon)^2 - |Z_i-Z_i'|^2}.
\end{align}
Using the trajectories defined by Eq.\ (\ref{rindlercoords1}) and (\ref{rindlercoords2}) with all other coordinates set to zero, the Wightman functions along a single accelerated trajectory are given by 
\begin{align}\label{wightmanrindler4}
    \mathcal{W}_R(s) &= - \frac{\kappa_M^2}{16\pi^2} \frac{1}{\sinh^2\big( \kappa_M s/2 - i\varepsilon \big)}. 
\end{align}
Since Eq.\ (\ref{wightmanrindler4}) is identical to Eq.\ (\ref{wightmandesitter}) after replacing $\kappa_M\Rightarrow \kappa$, we call these Wightman functions \textit{functionally equivalent}. Now, in the infinite-interaction time limit, the response function of an UdW detector on this trajectory is given by 
\begin{align}\label{rindlerresponse}
    \mathcal{F}_R^{\sigma\to\infty } (\Omega) &= \frac{\Omega}{2\pi } \frac{1}{e^{2\pi \Omega/\kappa_M} - 1}
\end{align}
which as expected, is identical to Eq.\ (\ref{staticresponse}) after replacing $\kappa_M \Rightarrow \kappa$. In such scenarios, we say that UdW detectors cannot operationally distinguish these spacetimes. This equivalence hints at the conformal relationship between the static de Sitter and Rindler metrics. To illustrate this, we review the coordinate transformation shown in \cite{Setare:2005cw}, which uses the static coordinate system $(t,r, \theta, \theta_B, \phi)$ to express the de Sitter metric in the form
\begin{align}
    \D s^2 &= - f(r) \D t^2 + \frac{\D r^2}{f(r)} -r^2 \D \Omega_3^2
\end{align}
where $\D \Omega_3^2$ is the line element for a 3-dimensional unit sphere in Euclidean space. One can then perform the coordinate transformation, 
\begin{align}
    \tau = l t, \qquad \xi = \frac{\sqrt{l^{-2} - r^2}}{\Omega}, \qquad Z_2 = \frac{r}{\Omega} \sin\theta\cos\theta_2,  \nonumber \\
    Z_3 = \frac{r}{\Omega}\sin\theta\sin\theta_2\cos\phi, \qquad Z_4 = \frac{r}{\Omega} \sin\theta\sin\theta_2\sin\phi 
\end{align}
where $\Omega = 1 - lr \cos\theta$. Under this transformation, the de Sitter line element takes on the form, 
\begin{align}
    \D s^2 &= -\Omega \big( \xi^2 \D \tau^2 - \D \xi^2 - \D Z_2^2 - \D Z_3^2 - \D Z_4^2\big) . 
\end{align}
Thus, we find that static de Sitter spacetime with line element $\D s^2_\text{dS}$ is conformally related to Rindler spacetime with line element $\D s^2_\text{R}$ \cite{candelasPhysRevD.19.2902}
\begin{align}
    \D s^2_\text{dS} &= \Omega \D s^2_\text{R},
\end{align}
where $\Omega$ is the conformal scaling factor. The conformal relationship between these metrics gives rise to effects such as Eq.\ (\ref{rindlerresponse}), whereby an UdW detector cannot operationally distinguish these spacetimes via a measurement of its response function. 

\section{Superpositions of angular separations}\label{sec:V}
We now apply the quantum-controlled UdW detector model to spatially translated superpositions in the static patch of de Sitter spacetime. The inclusion of the control degree of freedom allows us to model the detector as traveling in a quantum superposition of two spatially translated trajectories -- equivalently, 
to reside in a background spacetime in a quantum superposition of spatial translations. 

\begin{figure}[h]
    \centering
    \includegraphics[width=1\linewidth]{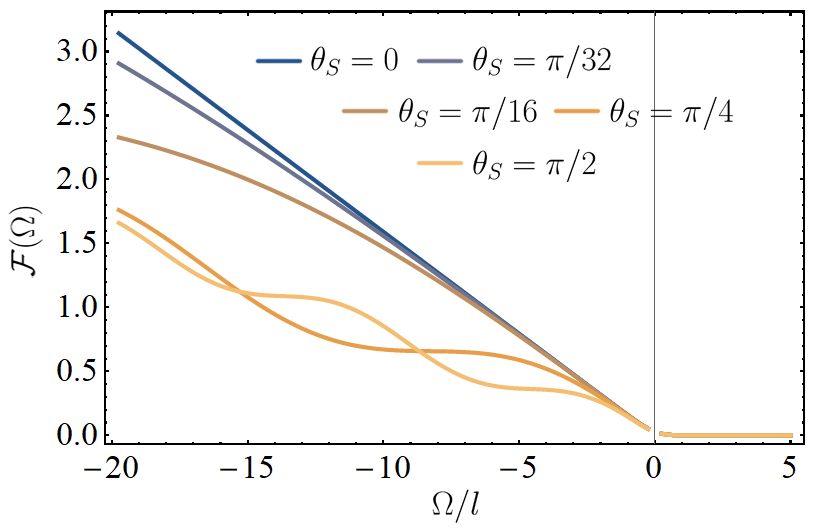}
    \caption{Response function of the detector superposed at different angular separations as a function of the energy gap, $\Omega/l$. We have fixed $R_D/l = 1/2$.}
    \label{fig:1response}
\end{figure}

We consider first  translations within the 5-dimensional Minkowski embedding space of the de Sitter spacetime (see Fig.\ \ref{fig:visualisation} for a visualisation). Note that this is not a limitation of our model; the translations occur within the embedding space according to our choice of the superposed coordinate $\theta$. Alternatively we could equally superpose the radial coordinate of the detector (which we do in Sec.~\ref{sec:VI}), which would be a translation on the de Sitter manifold itself. The trajectories associated with the quantum states in superposition differ by the angular separation $\theta_S = \theta_A - \theta_B$; these are parametrised by the coordinates 
\begin{align}
    \textbf{Z}^{(A)} &= \big( Z_0^{(A)} , Z_i^{(A)} \big) \\
    \textbf{Z}^{(B)} &= \big( Z_0^{(B)} , Z_i^{(B)} \big) ,
\end{align}
(the superscripts denoting the $i$th worldline of the superposition) where
\begin{widetext}
\begin{align}
\begin{split}
    Z_{0}^{ (A)} &= \sqrt{l^{-2} - R_D^2} \sinh (lt)  \\
    Z_{1}^{ (A)} &=  \sqrt{l^{-2} - R_D^2} \cosh (lt) \\
    Z_{2}^{ (A)}  &= R_D \cos\theta_A \vphantom{ \sqrt{l^{-2} -r^2}} \\
    Z_{3}^{ (A)} &= R_D \sin\theta_A \cos\phi \vphantom{ \sqrt{l^{-2} -R_D^2}} \\
    Z_{4}^{ (A)} &= R_D \sin\theta_A \sin\phi \vphantom{ \sqrt{l^{-2} -R_D^2}} 
\end{split}
\begin{split}
    Z_{0}^{ (B)} &= \sqrt{l^{-2} - R_D^2} \sinh (lt)  \\
    Z_{1}^{ (B)} &=  \sqrt{l^{-2} -R_D^2} \cosh (lt) \\
    Z_{2}^{ (B)} &= R_D \cos\theta_B \vphantom{ \sqrt{l^{-2} -r^2}} \\
    Z_{3}^{ (B)} &= R_D \sin\theta_B \cos\phi \vphantom{ \sqrt{l^{-2} -R_D^2}} \\
    Z_{4}^{ (B)} &= R_D \sin\theta_B \sin\phi \vphantom{ \sqrt{l^{-2} -R_D^2}}.
\end{split}
\end{align}
\end{widetext} 
This means that the superposed trajectories, each static at constant $(R_D,\theta_i,\phi)$, are translated (in the embedding space) by the Euclidean distance
\begin{align}\label{39a}
    \mathcal{L}_S &= 2 R_D \sin \left( \frac{\theta_S}{2} \right).    
\end{align}
We refer to $\mathcal{L}_S$ as the superposition distance. The translation occurs along an axis orthogonal to the radial direction from the horizon, so the individual amplitudes of the superposition experience equal cosmological redshifts and hence, equal temperatures, $\kappa\equiv \kappa_A = \kappa_B$. The translation takes on values in the finite domain $\mathcal{L}_S \in (0, 2/l)$. In the infinite interaction-time limit, the response function
\eqref{diffeopd} 
takes the form
\begin{widetext} 
\begin{align}\label{superposetheta72}
    \mathcal{F}_{\mathcal{L}_S}^{\sigma\to \infty } (\Omega) &= -\frac{\kappa^2}{32\pi^2} \infint\D s \:e^{-i\Omega s} \left\{ \frac{1}{\sinh^2 \big( \kappa s/2 - i\varepsilon\big)} + \frac{1}{\sinh^2 \big( \kappa s/2 - i\varepsilon \big) - \big( \kappa \mathcal{L}_S/2 \big)^2} \right\} .
\end{align}
The first integral of Eq.\ (\ref{superposetheta72}) gives the thermal response function of a single static trajectory while the second can be evaluated using the residue theorem (see Appendix \ref{appendixa}). Performing this calculation yields
\begin{align}\label{statictheta}
    \mathcal{F}_{\mathcal{L}_S}^{\sigma\to\infty}(\Omega) &= \frac{\Omega}{4\pi} \frac{1}{e^{2\pi\Omega/\kappa}-1} \left\{ 1 + \frac{\sin \big( 2 \Omega \kappa^{-1} \arcsinh \big( \mathcal{L}_S\kappa/2 \big) \big)}{\mathcal{L}_S \Omega \sqrt{1 + \big( \mathcal{L}_S\kappa/2 \big)^2}} \right\} .
\end{align}
\end{widetext} 
In Eq.\ (\ref{statictheta}), the response function has a local contribution from the individual trajectories of the superposition (yielding the familiar Planck distribution) and a nonlocal interference term between the trajectories. In the limit where the superposition distance vanishes, the interference term smoothly approaches the local term,
\begin{align}\label{eq77}
    \lim_{\mathcal{L}_S \to 0 }\mathcal{F}_{\mathcal{L}_S}^{\sigma\to\infty}(\Omega) &= \frac{\Omega}{2\pi} \frac{1}{e^{2\pi\Omega/\kappa} - 1}.
\end{align}
As $R_D$ approaches the cosmological horizon, the acceleration grows unbounded and the interference term vanishes. In this limit the Gibbons-Hawking temperature diverges, and the response function grows unbounded with the temperature of the field,
\begin{align}
  \lim_{R_D\to 1/l}\mathcal{F}^{\sigma\to \infty}_{\mathcal{L}_S}(\Omega)  &= \frac{\kappa}{8\pi^2}.
\end{align}
Even though the trajectories are separated by a finite Euclidean distance, any interference effects between them are washed out by the amplification of thermal particle production as the detector approaches the horizon. This leaves the response function in a classical mixture of contributions from the individual amplitudes of the superposition.
In Fig.\ \ref{fig:1response}, the detector response is plotted as a function of the energy gap, for various angular separations. For $\theta_S = 0$, we recover the Planckian spectrum of Eq.\ (\ref{eq77}), which grows linearly for increasing negative energy gaps (corresponding to a detector prepared in its excited state). As one introduces the spatial superposition, the response function oscillates with $\Omega/l$.
\begin{figure}[h]
    \centering
    \includegraphics[width=\linewidth]{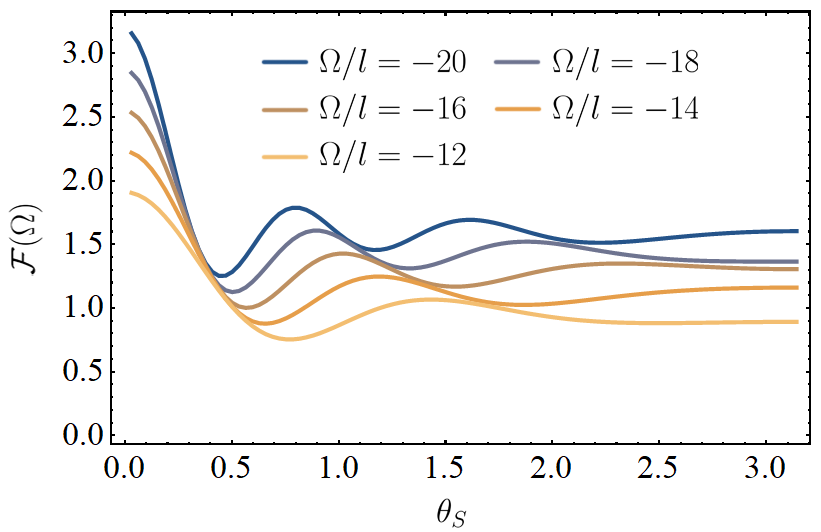}
    \caption{Response function of the detector with different negative energy gaps, as a function of the angular separation of the superposition, $\theta_S$. We have fixed $R_D/l = 1/2$. }
    \label{fig:2response}
\end{figure}
\begin{figure}[h]
    \centering
    \includegraphics[width=\linewidth]{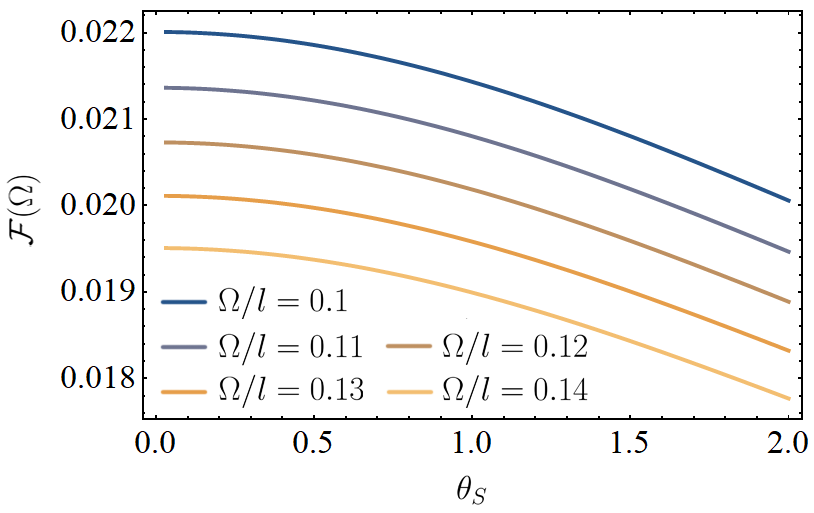}
    \caption{Response function of the detector with different positive energy gaps, as a function of the angular separation of the superposition, $\theta_S$. We have fixed $R_D/l = 1/2$. }
    \label{fig:3response}
\end{figure}
In Fig.\ \ref{fig:2response} and \ref{fig:3response}, we have plotted the detector response as a function of the angular separation of the superposition, for negative and positive energy gaps respectively. Since nonlocal correlations generally decrease with distance, we find likewise that the total response of the detector decays with the superposition distance.

The response function also satisfies the KMS detailed balance condition at temperature $T_\text{dS} = \kappa(2\pi)^{-1}$ for all $\mathcal{L}_S$,
\begin{align}
    \frac{\mathcal{F}_{\mathcal{L}_S}(\Omega)}{\mathcal{F}_{\mathcal{L}_S}(-\Omega)} = e^{-2\pi\Omega/\kappa}.
\end{align}
This contrasts with results found in \cite{foo2020unruhdewitt}, where a UdW detector was superposed along accelerated trajectories translated in the plane of motion, and the response was found in general to be nonthermal. This was due to   the asymmetric causal structure between the trajectories, leading to time-dependent interference effects that perturbed the quantum field away from thermalisation. For static worldlines in de Sitter spacetime, the causal symmetry of the trajectories nullifies these effects, allowing the detector to thermalise exactly. 

Interestingly, there is a functional equivalence between Eq.\ (\ref{statictheta}) and the response of a detector traveling in a superposition of accelerated trajectories in Minkowski spacetime, where the translation occurs along an axis orthogonal to the direction of motion. That is, for superposed worldlines in Minkowski spacetime defined by 
\begin{widetext} 
\begin{align}\label{42}
    \begin{split}
        Z_0^{ (A)} &= \kappa_M^{-1} \sinh(\kappa_M \tau) \\
        Z_1^{ (A)} &= \kappa_M^{-1} \cosh(\kappa_M \tau) \\
        Z_2^{ (A)} &= 0 \vphantom{\kappa_M^{-1}}
    \end{split}
    \begin{split}
        Z_0^{ (B)} &= \kappa_M^{-1} \sinh(\kappa_M\tau') \\
        Z_1^{ (B)} &= \kappa_M^{-1} \cosh(\kappa_M\tau') \\
        Z_2^{ (B)} &= \mathcal{L}_M \vphantom{\kappa_M^{-1}}
    \end{split}
\end{align}
and all other spatial coordinates equal to zero, the response function is given by (using similar techniques as the de Sitter case)
\begin{align}\label{frindler}
    \mathcal{F}^{\sigma\to\infty}_R(\Omega)&= \frac{\Omega}{4\pi}\frac{1}{e^{2\pi\Omega/\kappa_M} - 1} \left\{1 +  \frac{\sin \big( 2\Omega\kappa_M^{-1} \arcsinh \big( \mathcal{L}_M \kappa_M/2 \big) \big)}{\mathcal{L}_M\Omega \sqrt{1 + \big( \mathcal{L}_M\kappa_M/2 \big)^2}} \right\} . 
\end{align}
\end{widetext} 
The only difference between the response functions is that $\kappa_M$ is now the proper acceleration of the superposed detector trajectories in flat spacetime while $\mathcal{L}_M$ is still the Euclidean distance separating the trajectories. This means that the two spacetimes are operationally indistinguishable given the replacement $\kappa \Rightarrow \kappa_M$ and $\mathcal{L}_S \Rightarrow \mathcal{L}_M$. This correspondence reflects the conformal relationship between the Rindler wedge and static de Sitter spacetime, recalling that detectors on classical trajectories respond identically to the conformally coupled vacuum in static de Sitter space compared with uniformly accelerated detectors in the Minkowski vacuum. 

 The main difference between the two cases is that in Eq.\ (\ref{frindler}), the Euclidean distance between the accelerated trajectories, $\mathcal{L}_M$, takes on values on the full real line, so the interference vanishes when the trajectories are infinitely separated. This implies what we refer to as a \textit{quantitative equivalence} between the experience of detectors in these spacetimes. That is, for a particular subset of the $(l,R_D,\theta_S,\mathcal{L}_M)$ parameter space, the constraints
\begin{align}
    \mathcal{\kappa} &= \frac{1}{\sqrt{l^{-2}-R_D^2}} = \kappa_M \\
    \mathcal{L}_S &= 2R_D \sin\left( \frac{\theta_S}{2} \right) = \mathcal{L}_M
\end{align}
are satisfied exactly. In this case, the two spacetimes induce identical physical effects (i.e.\ the detected rate of particle production from the respective vacua) within UdW detectors. Such a correspondence can be considered to be a special case of the functional equivalence explained previously. A similar equivalence has been pointed out for the entanglement properties of two comoving detectors in de Sitter spacelike and two uniformly accelerating detectors
\cite{salton2015acceleration}.

\section{Superpositions of curvature}\label{sec:VI}

Next, we consider the detector traveling in a background spacetime in a quantum superposition of de Sitter curvatures. Our approach is purely phenomenological, in that our results are obtained by simply calculating the Wightman functions with respect to fields quantised on spacetime manifolds with two different (superposed) values of the de Sitter curvature. In our interpretation, there is still only one spacetime, now in a superposition of semiclassical parameters (i.e.\ the de Sitter curvature). Hence we expect our method for calculating detector observables in such spacetimes, which reduces to calculating the Wightman functions pulled back to the trajectories on these two manifolds, is not limited to scenarios involving a higher-dimensional embedding space. A conceptually helpful example is a black hole in a superposition of masses; even though the individual branches of the superposition correspond to unique solutions to Einstein's field equations, the detector still resides in a single spacetime, despite perceiving the Riemannian curvature to be superposed at each point in space. Finally, we do not posit the mechanism for how these kinds of superpositions are generated; however it is expected that a theory of quantum gravity should admit such solutions. 

\subsection{Equal temperatures}\label{seckms}
For comparison with the previous case, we first analyse the scenario where the two trajectories yield an identical KMS temperature, which allows analytic results to be obtained. This requires that $\kappa\equiv \kappa_A = \kappa_B$ with
\begin{align}\label{eq60}
    l_B &= \frac{l_A}{\sqrt{1 - l_A^2 \big( R_A - R_B \big) \big( R_A + R_B \big)}}
\end{align}
for static detectors at constant $(R_A,\theta_A,\phi)$ and $(R_B,\theta_B,\phi)$ in the respective states in the superposition. This further assumes that the radial coordinate $R_i$ of a given trajectory is perfectly correlated with the curvature $l_i$ along that trajectory. {The control system can be modelled as a maximally entangled two-qubit state,}
\begin{align}
    |\chi\rangle &= \frac{1}{\sqrt{2}} \big( |l_A\rangle |R_A \rangle + |l_B \rangle|R_B\rangle \big) 
\end{align}
where $|l_i\rangle$ are the curvature states and $|R_i\rangle$ are the radial states in superposition. One simply modifies the interaction Hamiltonian to read
\begin{align}
    \hat{H}_\text{int.} (\tau) &= \sum_{i=1}^N \hat{\mathcal{H}}_i \otimes |l_i \rangle \langle l_i | \otimes |R_i \rangle\langle R_i |.
\end{align}
Thus, the control prepares both the trajectory states of the detector and the curvature states of the spacetime. As before, it is straightforward to obtain the nonlocal Wightman functions for the detector, where the fields are now quantised on background metrics with different values of de Sitter curvature (and the detector travels along a trajectory with superposed radial coordinate). The response function  
\eqref{diffeopd}  in the infinite interaction time limit is given by 
\begin{widetext} 
\begin{align}\label{superpositionofcurvatures}
    \mathcal{F}_{R_D,\: l}^{\sigma\to \infty} (\Omega) &= \frac{\Omega}{4\pi } \frac{1}{e^{2\pi\Omega/\kappa} - 1}\left\{ 1 + \frac{\kappa \sin\big( 2\Omega\kappa^{-1} \arcsinh  \big(\zeta_{R_D,\: l} \big) \big) }{2\zeta_{R_D,\: l}\Omega\sqrt{1 + \zeta_{R_D,\: l}^2}}  \right\} 
\end{align}
\end{widetext}
where the interference term in the bracket takes the same functional form as the previous cases, and we have defined
\begin{align}
    \zeta_{R_D,\: l}^2 &= \frac{\kappa^2}{4} \left( \frac{1}{l_A^2} + \frac{1}{l_B^2} \right) - \frac{1}{2} \left(1+  \kappa^2 R_AR_B \cos \theta_S \right)
\end{align}
\begin{figure}[h]
    \centering
    \includegraphics[width=\linewidth]{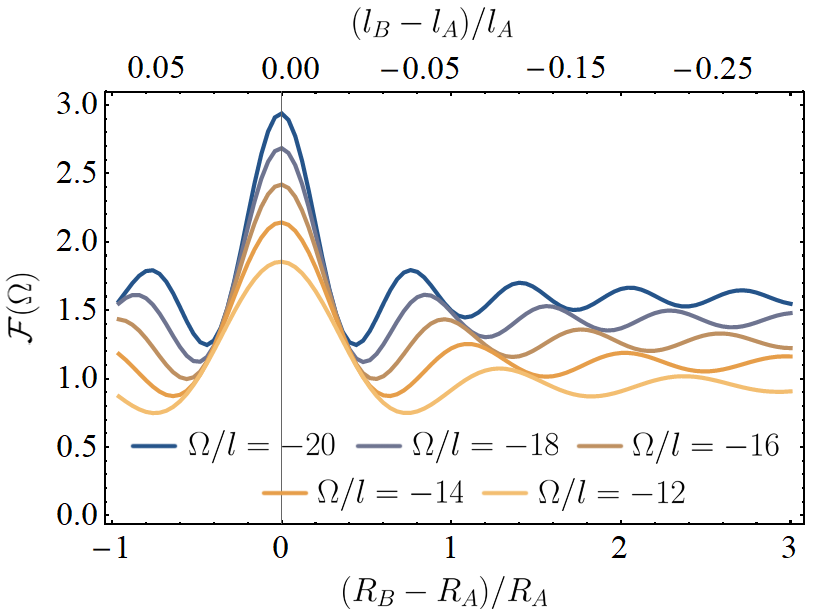}
    \caption{Response function of the detector as a function of the radial separation (curvature separation) of the superposition, for different negative energy gaps. We have fixed $R_A/l_A = 1$ and $\theta_S = \pi/32$. }
    \label{fig:4response}
\end{figure}
\begin{figure}[h]
    \centering
    \includegraphics[width=\linewidth]{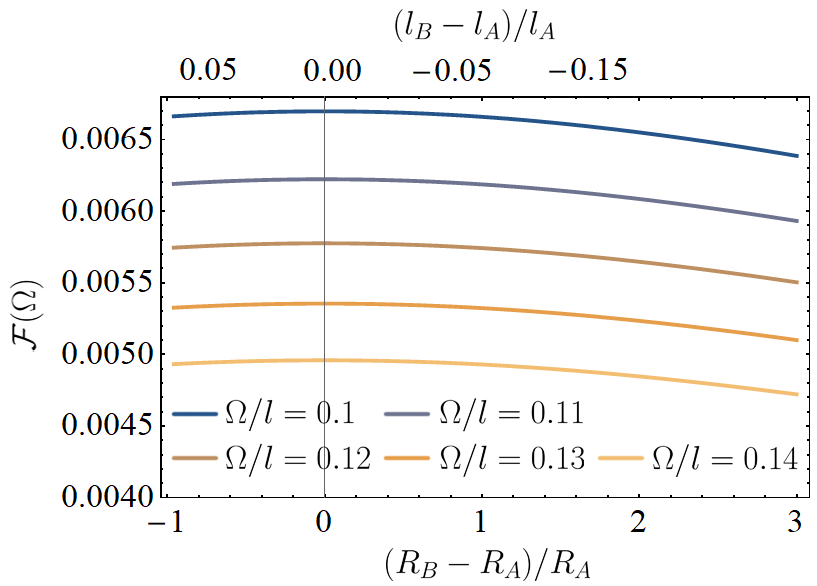}
    \caption{Response function of the detector as a function of the radial separation (curvature separation) of the superposition, for different positive energy gaps. We have fixed $R_A/l_A = 1$ and $\theta_S = \pi/32$.}
    \label{fig:5response}
\end{figure}
As with the prior two cases, the detector response function satisfies the detailed balanced criterion for all parameter values at the tempreature $T_\text{dS} = \kappa(2\pi)^{-1}$. This is plotted in Fig.\ \ref{fig:4response} and \ref{fig:5response}, showing the detector response for negative and positive energy gaps as a function of the radial separation of the trajectories in superposition (this radial separation is correlated with the difference in the de Sitter lengths in the respective branches of the superposition, also shown). Similar to the angular superposition, the response function decays in an (non-)oscillatory manner for (positive) negative energy gaps as the spatial separation between the branches increases.

\begin{figure}[h]
    \centering
    \includegraphics[width=\linewidth]{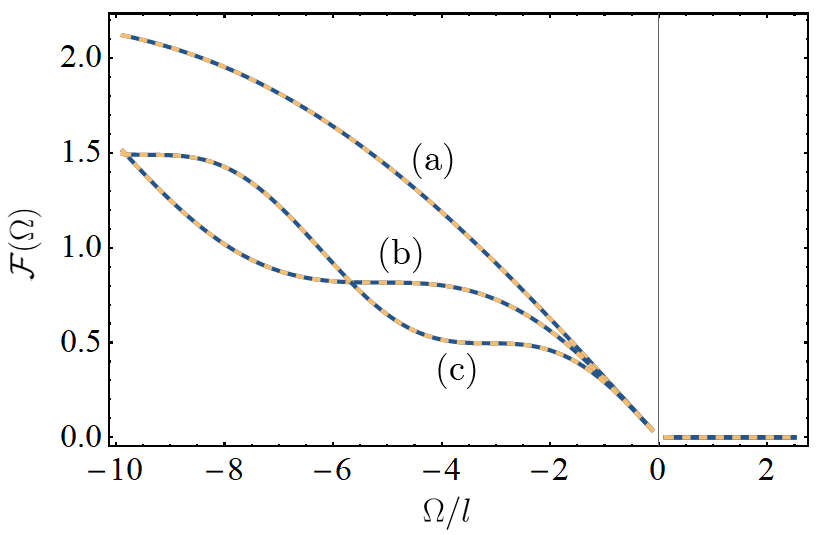}
    \caption{Response function of the detector as a function of the energy gap, for (blue) superpositions of angular coordinates and (yellow) an equal temperature superposition of curvatures. For the superposition of curvatures, we have used (a) $R_A/R_B = 0.83$, (b) $R_A/R_B = 0.625$ and (c) $R_A/R_B = 0.5$ and the corresponding angular superposition yielding a quantitative equivalence. }
    \label{fig:7response}
\end{figure}

We make several observations. First, the response function Eq.\ (\ref{superpositionofcurvatures}) is functionally equivalent and hence operationally indistinguishable from Eq.\ (\ref{statictheta}) for the spacetime in a superposition of angular variables, after the association $\zeta_{R_D,\: l} \Rightarrow \mathcal{L}_S \kappa/2$; indeed these quantities are equal if $l_A=l_B$. One can also demonstrate the quantitative equivalence of these two scenarios for a specific subset of the parameter space. Firstly, we require that the temperature of the field perceived by the detectors is identical,
\begin{align}
    \underbrace{\frac{1}{\sqrt{l^{-2}-R_D^2}}}_\text{$\theta$ superposition} &\equiv \underbrace{\frac{1}{\sqrt{l_i^{-2} - R_i^2}}}_\text{$R_D$, $l$, $\theta$ superposition}
\end{align}
 which constrains the possible values of $l$, $l_i$ and $R_D$, $R_i$. If for example, one takes $l = l_A$ and $R_D  = R_A$, and that $\theta_A = \theta_B$ in the curvature superposition, then the two scenarios are quantitatively equivalent when
\begin{align}
    \theta_{S} &= 2 \sin^{-1} \bigg\{ \frac{R_A - R_B}{2R_A} \bigg\}
\end{align}
where $\theta_S$ is the angular separation of Eq.\ (\ref{39}) -- i.e.\ when $(r,l)$ have classical values. Therefore a UdW detector traveling on a specific subset of superposed static trajectories -- trajectories in a superposition of angular separations $\theta_S$ -- responds to the quantum field identically to certain cases where its worldline is defined by a superposition of radial distances from the cosmological horizon, with the spacetime possessing quantised curvature degrees of freedom. In Fig.\ \ref{fig:7response}, we have plotted the response of the detector in the two scenarios already considered (superposition of angular coordinates, and equal temperature superpositions of curvature), for parameters which yield an identical response in the detector (i.e.\ demonstrating examples of the quantitative equivalence between the cases).

Both the functional and quantitative correspondences between the two de Sitter spacetimes in superposition extend to the spatially translated Rindler trajectories in Eq.\ (\ref{42}). Clearly, if one associates $\kappa \Rightarrow \kappa_M$ and $\zeta_\text{dS$-r,$} \Rightarrow \mathcal{L}_M\kappa_M/2$, the response functions Eq.\ (\ref{frindler}) and Eq.\ (\ref{superpositionofcurvatures}) are identical. Moreover, there exists some region in the $(l_i, R_i, \theta_i)$ parameter space for which $\kappa = \kappa_M$ and $\zeta_{R_D,\: l} = \mathcal{L}_M\kappa_M/2$ exactly, indicating a quantitative equivalence between the two cases. Table \ref{ref:table1} summarises the equivalences between these cases.
\begin{table}[h!]
\begin{center}
\begin{tabular}{c c c}
superposition of orthogonally \\
translated Rindler trajectories \\\\
$\Updownarrow$ \\\\
superposition of translated equal-redshift \\
trajectories in de Sitter \\\\
$\Updownarrow$ \\\\
superposition of curvatures + detector\\
radial coordinates w. equal temperatures in de Sitter
\end{tabular}
\caption{Operational equivalence of the superposition of translated Rindler trajectories, superpositions of spatial translations in de Sitter, and equal-temperature curvature superpositions. }
\label{ref:table1}
\end{center}
\end{table}

The operational equivalence between the de Sitter spacetimes in superposition and superpositions of accelerated trajectories in Minkowski space is linked to the conformal relationship between the Rindler and static de Sitter geometries, discussed previously. The fact that the two-point correlation functions -- which characterise the experience of UdW detectors -- are effectively identical in these radically different scenarios is not wholly surprising. Beyond this, one can recall the Einstein Equivalence Principle (EEP), which generally asserts that certain classes of noninertial motion elicit the same physical effects as gravitational fields. Here, we can apply the EEP beyond classical trajectory states; the detection of field quanta in the Minkowski vacuum according to detectors in a superposition of trajectory states can be mapped to a comparable scenario on a curved de Sitter spacetime in superposition. We conjecture that other classes of noninertial motion in superposition may possess operational correspondences with different spacetime metrics in superposition.

A final observation is that Eq.\ (\ref{superpositionofcurvatures}) smoothly approaches Eq.\ (\ref{statictheta}) in the limit where the superposition of curvatures vanishes
\begin{align}
\lim_{l_A\to l_B} \mathcal{F}_{R_D,\: l}^{\sigma\to \infty }(\Omega) &= \mathcal{F}_{\mathcal{L}_S}^{\sigma\to\infty } (\Omega).
\end{align}
This result is interesting because one should not necessarily expect \textit{a priori} that this must be the case. On one hand, the detector-spacetime system in a superposition of spatial translations is a relatively innocuous setup, since it can be understood as a superposition of trajectory states on a classical spacetime manifold. Meanwhile the present case, which includes superpositions of curvature states, does not possess such an analogy. This is because the fields $\hat{\Phi}( \mathsf{x}_i(\tau ) )$ and $\hat{\Phi}( \mathsf{x}_j(\tau) )$ are quantised on spacetime manifolds which individually represent unique solutions of general relativity. Hence, one cannot perform \textit{any} global coordinate transformation that reduces the detector's interaction with these fields to a superposition of two trajectories but on a single metric with one, global value for the curvature. 

\subsection{Unequal temperatures}
We conclude by studying the detector in static de Sitter spacetime in a superpositon of curvatures, without requiring that the local temperatures associated with the superposed amplitudes be equal.  Unlike the prior cases, it can be straightforwardly shown that the detector will not thermalise at the temperature associated with either of the amplitudes in superposition. More specifically, the response function of the detector in an $N$-amplitude superposition, where each amplitude has an associated temperature, is
\begin{align}
    \mathcal{F}(\Omega) &= \sum_{i,j}^N \mathcal{F}_{ij}(\Omega) = \sum_{i = j}^N \mathcal{F}_{ij}(\Omega) + \underbrace{\sum_{i\neq j} \mathcal{F}_{ij}(\Omega)}_{\mathcal{F}^\text{int.}(\Omega)}.
\end{align}
For $N=2$, the excitation-to-deexcitation ratio for the response function is given by 
\begin{align}
    \frac{\mathcal{F}(\Omega)}{\mathcal{F}(-\Omega)} &= \frac{\mathcal{F}_A(\Omega) + \mathcal{F}_B(\Omega) + \mathbin{F}^\text{int.}(\Omega)}{\mathcal{F}_A(-\Omega) + \mathcal{F}_B(-\Omega) + \mathcal{F}^\text{int.}(-\Omega)} \\ 
    &\neq \frac{\mathcal{F}_A(\Omega)}{\mathcal{F}_A(-\Omega)} , \frac{\mathcal{F}_B(\Omega)}{\mathcal{F}_B(-\Omega)}
\end{align}
where we have used the shorthand $\mathcal{F}_i(\Omega)$ where $i = A,B$ to denote the two paths in superposition. This result, namely that the superposition of two thermal channels at different temperatures enacted upon an UdW detector does not lead to its thermalisation, was shown in \cite{foo2020unruhdewitt} for a detector in a superposition of proper accelerations in flat spacetime. Returning to the present scenario, we consider the superposition of detector trajectories defined by the parametrisation
\begin{widetext} 
\begin{align}
\begin{split}
    Z_0^{ (A)} &= \kappa_A^{-1} \sinh(\kappa_A \tau) \\
    Z_1^{ (A)} &= \kappa_A^{-1} \cosh(\kappa_A \tau ) \\
    Z_2^{ (A)} &= R_A\cos\theta_A \vphantom{ \kappa_A^{-1} } \\
    Z_3^{ (A)} &= R_A\sin\theta_A \cos\phi \vphantom{ \kappa_A^{-1} } \\
    Z_4^{ (A)} &= R_A\sin\theta_A \sin\phi \vphantom{ \kappa_A^{-1} }
\end{split}
\begin{split}
    Z_0^{ (B)} &= \kappa_B^{-1} \sinh(\kappa_B\tau) \\
    Z_1^{ (B)} &= \kappa_B^{-1} \cosh(\kappa_B\tau) \\
    Z_2^{ (B)} &= R_B\cos\theta_B \vphantom{ \kappa_A^{-1} } \\
    Z_3^{ (B)} &= R_B\sin\theta_B \cos\phi \vphantom{ \kappa_A^{-1} } \\
    Z_4^{ (B)} &= R_B \sin\theta_B \sin\phi  \vphantom{ \kappa_A^{-1} }
\end{split}
\end{align}
\end{widetext} 
where as usual $\kappa_i = \big( l_i^{-2} - R_i^2 \big)^{-1/2}$. The nonlocal Wightman functions become
\begin{align}\label{diffkappa}
    \mathcal{W}_{R_D,\:l}\big(\mathsf{x}_A,\mathsf{x}_B'\big) &= - \frac{\kappa_A\kappa_B}{16\pi^2}\frac{1}{\sinh^2 \big( w /2 - i\varepsilon \big) - \gamma_{R_D,\:l}^2 }
\end{align}
which are now time-dependent, having defined $w = \kappa_A \tau - \kappa_B (\tau-s)$ and
\begin{align}
    \gamma_{R_D,\:l}^2 &= \frac{\kappa_A\kappa_B}{4} \left( \frac{1}{l_A^2} + \frac{1}{l_B^2} \right) - \frac{1 + \kappa_A \kappa_B R_A R_B \cos \theta_S }{2} .
\end{align}
Equation (\ref{diffkappa}) is that used to obtain the response function of Eq.\ (\ref{superpositionofcurvatures}) without assuming equal KMS temperatures on the individual amplitudes of the superposition. 

\begin{figure}[h]
    \centering
    \includegraphics[width=\linewidth]{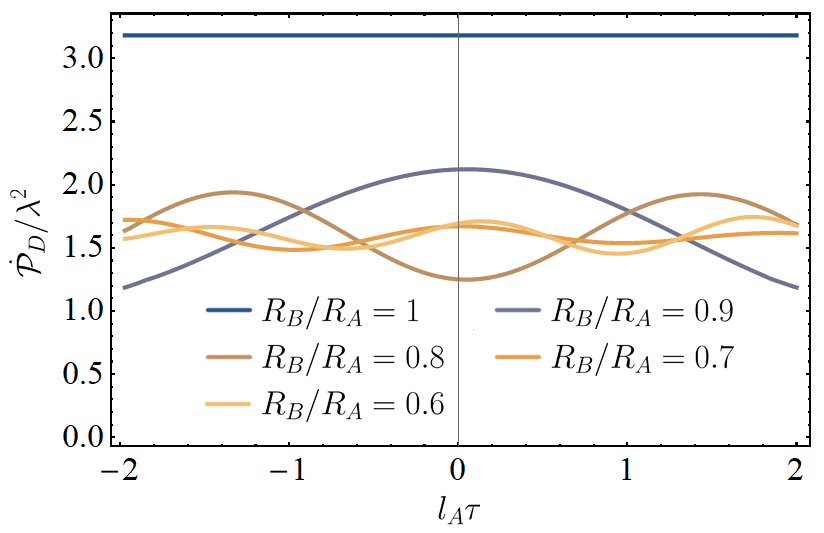}
    \caption{Transition rate for the detector in a superposition of radial coordinates as a function of the proper time, for different radial separations. We have fixed $\Omega/l_A = -40$. }
    \label{fig:9rate}
\end{figure}

There are two special cases which take on radically different physical interpretations. The first is for a fixed background spacetime with the detector in a superposition of radial locations from the horizon, $l \equiv l_A=l_B$, so that
\begin{align}\label{68}
    \gamma_{R_D}^2 &=  \frac{\kappa_{RA}\kappa_{RB}}{2l^2} - \frac{1}{2} \big( 1 + \kappa_{RA}\kappa_{RB} R_A R_B\cos\theta_{RS} \big)
\end{align}
where $\kappa_{Ri} = \big( l^{-2} - R_i^2\big)^{-1/2}$ for $i = A, B$ and $\theta_{RS}$ is the angular separation of the superposition. In this case, the control state only affects the trajectories of the detector. As before, there exists some unitary transformation that maps the superposed trajectory states to a single classical trajectory, so that the detector-field dynamics undergo a modified evolution in which the fields are quantised on a background in a superposition of spacetime spatial states. The two scenarios, as argued previously, are diffeomorphic invariant.
\begin{figure}[h]
    \centering
    \includegraphics[width=\linewidth]{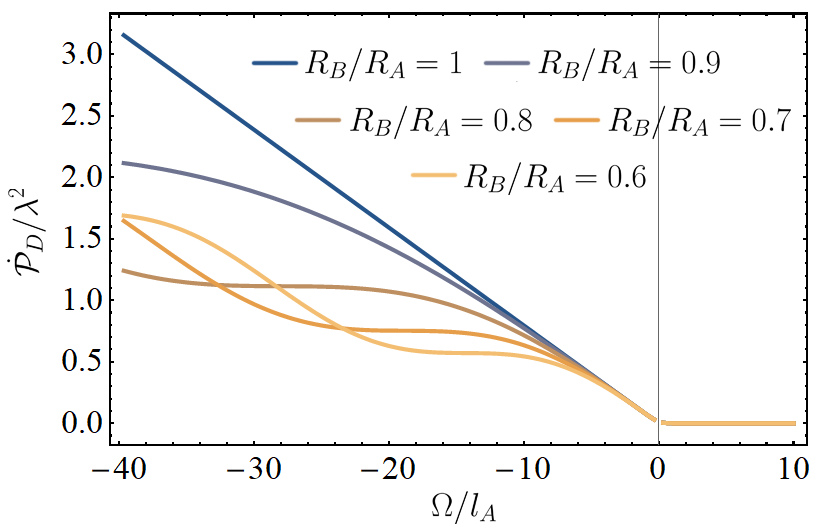}
    \caption{Transition rate for the detector in a superposition of radial coordinates as a function of the energy gap, at the fixed proper time $l_A \tau = 0$. We have also fixed $R_A/l_A = 2$. }
    \label{fig:8rate}
\end{figure}
In Fig.\ \ref{fig:8rate}, we have plotted the transition rate as a function of the energy gap at the timeslice $l_A \tau = 0$. The transition rate profile resembles that of the detector response when the individual branches of the superposition had equal redshifts. Meanwhile in Fig.\ \ref{fig:9rate}, the transition rate is plotted as a function of the proper time of the detector for different radial separations. For detectors on classical paths, the transition rate is time-independent; the introduction of the superposition elicits time-dependent oscillations.

In the second case, we take the background spacetime to be in a superposition of curvatures, with the detector on a single static trajectory, $R_D\equiv R_A=R_B$. This yields,
\begin{align}\label{71}
    \gamma_{l}^2 &= \frac{\kappa_{lA}\kappa_{lB}}{4} \left( \frac{1}{l_A^2} + \frac{1}{l_B^2} \right) - \frac{1}{2}\big( 1 + \kappa_{lA}\kappa_{lB} R_D^2\cos \theta_{lS}  \big)
\end{align}
where $\kappa_{li} = \big( l_i^{-2} - R_D^2 \big)^{-1/2}$ for $i = A,B$ and $\theta_{lS}$ is the angular separation of the superposition. In this case, the control system solely affects  the quantum state of the spacetime, defining its Hilbert space and controlling the superposed values of the curvature. Unlike superpositions of detector trajectory states, here there is no diffeomorphism which maps the individual amplitudes of the superposition to a single, classical spacetime background. 

\begin{figure}[h]
    \centering
    \includegraphics[width=\linewidth]{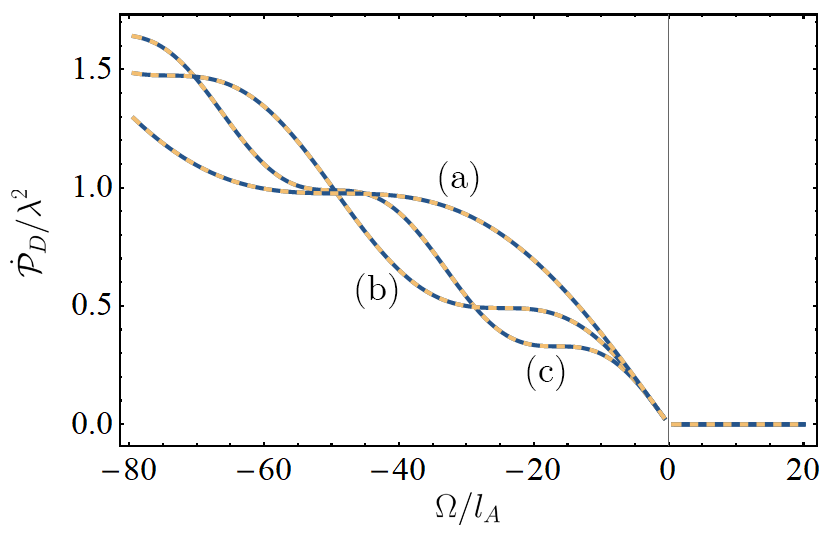}
    \caption{Transition rate of the detector as a function of the energy gap for (blue) superpositions of detector radii and (yellow) superpositions of curvature. We have shown the transition rate for (a) $R_B/R_A = 0.25$, (b) $R_B/R_A = 0.5$ and (c) $R_B /R_A = 0.75$, with the corresponding ratios of $l_B/l_A$ for which a quantitative equivalence between the two scenarios exists. }
    \label{fig:10rate}
\end{figure}

Although the Wightman functions only differ by an additive constant in the denominator, Eq.\ (\ref{68}) and (\ref{71}), and can hence be understood as functionally equivalent, it is not immediately clear whether a quantitative equivalence between the two cases exists. This is because the parameters $(l,l_i)$ and $(R_D,R_i)$ are additionally constrained by the requirement that the surface gravities associated with the superposed amplitudes of the spacetime need to be equal; $\kappa_{Ri} = \kappa_{li}$. The question then arises if there is a subset of the parameter space for which the nonlocal Wightman functions, and hence the response of the detector, is identical between these two cases? We find that the condition
\begin{align}\label{90condition}
    \gamma_{R_D} = \gamma_{l}
\end{align}
is only satisfied when $l = l_A$ or $l_B$ \textit{and} $R_D = R_A$ or $R_B$. In such a circumstance (taking for example $l = l_A$ and $R_D = R_A$) then a further constraint on the angular separation is required in order for Eq.\ (\ref{90condition}) to hold:
\begin{align}\label{condition1}
    \theta_{lS} &= \pm \cos^{-1}  \bigg\{ \frac{1}{2} \bigg( \frac{R_A^2 - R_B^2}{R_A} + 2 R_B \cos \theta_{RS} \bigg) \bigg\}
\end{align}
or in the special case where $\theta_{RS} = \theta_{lS}$, 
\begin{align}\label{condition2}
    \theta_{RS} = \theta_{lS} &= \pm \cos^{-1} \bigg( \frac{R_A^2 - R_B^2}{2R_A \big(1 - R_B \big)}  \bigg).
\end{align}
Thus, under these constraints, the two scenarios can be understood to be operationally equivalent according to UdW detectors. Indeed in Fig.\ \ref{fig:10rate}, we have plotted the transition rate using parameters which yield such an equivalence; i.e.\ where a superposition of radial coordinates induces the same detector statistics as a detector embedded in a spacetime with superposed de Sitter curvatures. Outside this narrow parameter space, the spacetimes could in principle be distinguished from each other via measurements of the transition rate.
Of course, we are interested in this distinguishability not for practical but rather fundamental reasons. One might intuitively expect that genuine superpositions of spacetime geometries (conceivably encapsulated within a theory of quantum gravity) would yield predictions without any such analogy. Apart from the special cases of Eq.\ (\ref{condition1}) and (\ref{condition2}), a detector in a superposition of radial coordinates can distinguish such a scenario from a de Sitter spacetime in a superposition of curvature states.

Based on the prior correspondence found for superpositions of spatial translations in Rindler space and de Sitter space, we might expect an operational equivalence between the present scenarios and a detector traveling in a superposition of proper accelerations. In particular, for a detector in flat spacetime with superposed trajectories parametrised by 
\begin{widetext} 
\begin{align}
\begin{split}
    Z_0^{ (A)} &= \kappa_{MA}^{-1} \sinh( \kappa_{MA} \tau ) \\
    Z_1^{ (A)} &= \kappa_{MA}^{-1} \cosh ( \kappa_{MA} \tau) \\
    Z_i^{ (A)} &= 0 \vphantom{\kappa_{MA}^{-1}}
\end{split}
\begin{split}
    Z_0^{ (B)} &= \kappa_{MB}^{-1} \sinh ( \kappa_{MB} \tau ) \\
    Z_1^{ (B)} &= \kappa_{MB}^{-1} \cosh(\kappa_{MB} \tau ) \\
    Z_i^{ (B)} &= 0 \vphantom{\kappa_{MB}^{-1}}
\end{split}
\end{align}
\end{widetext} 
where $\kappa_{Mi}$ are the proper accelerations of the two trajectories, the nonlocal Wightman functions take on the form
\begin{align}\label{diffacc105}
    \mathcal{W}_{R}\big(\mathsf{x}_A , \mathsf{x}_B' \big) &=  - \frac{\kappa_{MA}\kappa_{MB}}{16\pi^2} \frac{1}{\sinh^2 \big( w_M/2 - i\varepsilon \big) - \gamma_R^2} 
\end{align}
where $w_M = \kappa_{MA}\tau - \kappa_{MB}(\tau - s ) $ and we have defined
\begin{align}
    \gamma^2_R &= \frac{\kappa_{MA}^2 + \kappa_{MB}^2}{4\kappa_{MA}\kappa_{MB}} - \frac{1}{2}.
\end{align}
The functional equivalence of the nonlocal Wightman functions -- compare Eq.\ (\ref{diffkappa}) with Eq.\ (\ref{diffacc105}) -- is quite clear, with the association $\kappa_i \Rightarrow \kappa_{Mi}$ and $\gamma_{R_D,\:l}\Rightarrow \gamma_R$. However a quantitative equivalence between the two scenarios only exists when (in the de Sitter superposition) $R_A = R_B = 0$. Only in this special case does the detector on a de Sitter manifold in a superposition of curvatures respond (in the quantitative sense) identically to the quantum field as it would in flat spacetime when traveling in a superposition of proper accelerations, given that $l_i^{-1} = \kappa_{Mi}$. 


\section{Probing spacetime superpositions with UdW detectors}\label{sec:VII}

Our results have shown that discerning a genuine superposition of spacetime metrics from superpositions of trajectory states in flat spacetime -- at least with idealised quantum probes characterised by the UdW model -- reduces to the equivalence or non-equivalence of the nonlocal Wightman functions that encode the field correlations between the quantum states in superposition. Of course in the previous analyses, we have studied cases wherein a meaningful analogy (through application of the conformal invariance of the spacetimes and the EEP) between the classes of superposition states (spacetime superpositions compared with certain trajectory superpositions in flat spacetime) can be made. Our results are not merely interesting because de Sitter spacetime in superposition is operationally equivalent to certain classes of Rindler trajectories in superposition, but that any such analogy exists at all. This is because quantum states of spacetime are considered beyond the reach of current physical theories, and some postulate they cannot even in principle be described \cite{carlip2008quantum,Penrose:1996cv}. 

To reiterate our point, let us qualitatively consider a final example. In particular, we compare the behaviour of a detector superposed along static trajectories in a finite-temperature thermal bath in Minkowski spacetime with a static detector outside a (1+1)-dimensional Schwarszchild black hole, superposed at two radial distance from the detector. For comparison with known results \cite{Tjoa:2020eqh}, we will assume the detectors are coupled to the derivative of the massless scalar field. The single-trajectory derivative Wightman functions, $\mathcal{A}_i (\mathsf{x}_i, \mathsf{x}_i' )$, are respectively \cite{weldon2000thermal}
\begin{align}
    \mathcal{A}_T \big( \mathsf{x}_i , \mathsf{x}_i' \big) &= - \frac{\kappa^2}{8\pi} \frac{1}{\sinh^2( \kappa s/2 - i\varepsilon)} 
\end{align}
for the detector immersed in the thermal bath at temperature $\kappa(2\pi)^{-1}$, and
\begin{align}\label{68:2}
    \mathcal{A}_H\big( \mathsf{x}_i, \mathsf{x}_i' \big)  &= - \frac{1}{8\pi} \frac{(4 \sqrt{f(R_D)}M)^{-2}}{\sinh^2( s/( 8 \sqrt{f(R_D)}M) - i\varepsilon ) }
\end{align}
for the detector outside the black hole \cite{Tjoa:2020eqh}, where $M$ is the black hole mass, $f = 1 - 2M/R_D$ is the usual metric function for the Schwarzschild spacetime, and $R_D$ is the radial distance from the black hole. In Eq.\ (\ref{68:2}), the subscript $H$ denotes that the Wightman function is evaluated with respect to the Hartle-Hawking-Israel vacuum, which represents a black hole at thermal equilibrium with a radiation bath at the Hawking temperature, $T_H = ( 8\pi M)^{-1}$. Clearly the two Wightman functions are functionally equivalent after making the association $\kappa \Rightarrow (4\sqrt{f} M)^{-1}$. We can refer to these situations as operationally indistinguishable. 

Below, we analyse whether a detector can discern if it is in a spatial superpostion 
in the flat spacetime thermal bath, or if it resides in a metric generated by a superposition of black hole spatial locations. Let us assume that the black hole is in a superposition of radial distances from the detector (equivalently, the detector is prepared in a superposition of radial locations outside the black hole). For a detector far from the superposed black hole horizon $(R_D \gg 2M)$, the Wightman functions approach those for superposed trajectories in thermal Minkowski spacetime 
\begin{align}
    \lim_{R_D\gg 2M} \mathcal{A}_H\big(\mathsf{x}_i, \mathsf{x}_j' \big) = \mathcal{A}_T\big( \mathsf{x}_i, \mathsf{x}_j'\big) 
\end{align}
where
\begin{align}\label{80}
    \mathcal{A}_T\big( \mathsf{x}_i , \mathsf{x}_j' \big) =  - \frac{1}{4\pi}\frac{\text{csch} \big( \frac{R_S - s + i\varepsilon}{8M} \big) + \text{csch} \big( \frac{R_S + s - i\varepsilon }{8M} \big) }{64M^2} .
\end{align}
In Schwarzschild, $R_S = R_A - R_B$ is the radial distance over which the black hole is superposed, while in Minkowski it represents the fixed distance between the superposed static trajectories. In this limit, the local Tolman temperatures \cite{TolmanPhysRev.35.904},
\begin{align}
    T_T &= \big( 8 \pi M \sqrt{1 - 2M / R_D } \big)^{-1} \simeq \big( 8 \pi M \big)^{-1}
\end{align}
perceived by the detector on each amplitude of the superposition are approximately equal to the Hawking temperature of the black hole, yielding the stationary Wightman functions, Eq.\ (\ref{80}). Thus, the detector in the black-hole-superposition spacetime is immersed in an approximately thermal bath, directly corresponding to the flat spacetime thermal bath analogy. 

However in the near-horizon limit, the Schwarzschild Wightman functions become 
\begin{align}\label{71:2}
    \mathcal{A}_H \big( \mathsf{x}_i , \mathsf{x}_j' \big) &= - \frac{1}{4\pi} \frac{1}{64M^2} \frac{1}{\sqrt{f_A f_B}} \nonumber \\ 
    & \times \bigg\{ \text{csch}^2 \Big( \frac{R_S - w_N}{8M} \Big) + \text{csch}^2\Big( \frac{R_S + w_N}{8M} \Big) \bigg\}
\end{align}
where $w_N = \tau/\sqrt{f_A} - (\tau - s ) / \sqrt{f_B} - i\varepsilon$ and $f_i = 1 - 2M/R_i$. Equation (\ref{71:2}) is no longer stationary, since the detector will experience unequal redshifts produced by the superposed amplitudes of the black holes radial position. Thus, in such a scenario, one can operationally distinguish a flat spacetime case of a detector in superposition of trajectories from a black hole in a genuine superposition of position states (and hence, one which generates a background metric possessing a quantised value of curvature throughout the entire spacetime). This equivalence breaking is due to the fact that in the near-horizon limit, the detector is effectively interacting with thermal channels at different temperatures, in a coherent superposition.  On the other hand, one could argue that the comparison is no longer meaningful since the two scenarios are fundamentally different.

Finally, we note that the near-horizon regime is exactly that in which two detectors on classical trajectories, through the correlations they harvest from the quantum field (in a process known as entanglement harvesting \cite{salton2015acceleration,ver2009entangling,Tjoa:2020eqh,Martin-Martinez:2015qwa}), can differentiate the spacetimes. The addition of a second detector allows one to access the nonlocal correlations between the classical trajectories that they traverse (see Appendix \ref{entanglementharvesting}); these are exactly the correlations already accessible to single detectors in superposition. Implicitly we are only addressing spatial superpositions of metrics, since, as we have been emphasising, a superposition of mass or curvature states has no classical analog. That is, it is not physically meaningful to talk about two classical detectors, each residing in its own universe with a different global curvature, for example (in our model, this is just a single spacetime in quantum superposition). In sum, we infer that a single UdW detector can discern a spatial superposition of spacetime metrics in regimes where entanglement harvesting between two detectors can achieve an analogous task.

\section{Conclusion}\label{sec:VIII}
The main purpose of this paper has been to address foundational questions surrounding the phenomenology of superpositions of spacetime metrics. We have approached the problem from the bottom-up, namely through an operational model using Unruh-deWitt detectors.  In particular, we considered the experience of such detectors when situated in a static de Sitter spacetime in a superposition of spatial (i.e.\ angular separations in the embedding space as well as radial superpositions) and curvature states. In the former example, we showed that there always exists a general coordinate transformation which maps the manifold to a  classical one while mapping the detector trajectory to a superposition state, and vice versa. For spacetimes in a superposition of curvatures, such a diffeomorphism does not exist. This is because the gravitational source of curvature associated with the individual amplitudes of the superposition constitutes a unique solution of Einstein's field equations; no diffeomorphism mapping between such solutions exists. 

We found that for particular superpositions of the spacetime, even those in which the de Sitter curvature 
is quantised and are thus not diffemoporhic, an operational correspondence nevertheless exists with certain classes of superposed trajectory states of uniformly accelerated detectors in the Minkowski vacuum. This correspondence can be understood from two fronts; the conformal relationship between the Rindler and de Sitter geometries, as well as a generalisation of the EEP, which relates the physical effects induced by classes of noninertial motion with analogous gravitational settings. We also argued that the regimes in which UdW-like detectors can differentiate between genuine quantum superpositions of spacetime metrics and mere superposition states of detector motion are exactly those in which two detectors on classical worldlines (corresponding to the superposed worldlines of the quantum-controlled detector) can achieve such a task. 

As noted, our model for studying the operational effects of quantum spacetimes is not limited to those embedded within a higher-dimensional space; one could conceivably apply our approach to, say, a black hole in a superposition of masses or angular rotations. An important question to address in future investigations are the limitations and constraints of our new approach. For example, it is unclear whether one can effectively model superpositions of \textit{different} spacetime metrics, such as de Sitter spacetime superposed with Schwarzschild spacetime. Issues such as the calculation of mode-sum expansions and defining a vacuum state highlight this. Hence, there may be something akin to superselection rules for spacetime superpositions.

\section{Acknowledgments} J.F thanks Zehua Tian for helpful discussions regarding the complex integrals shown in this paper. M.Z. acknowledges support from DECRA Grant DE180101443. R.B.M acknowledges support from the Natural Sciences and Engineering Research Council of Canada and from  AOARD Grant FA2386-19-1-4077.

\appendix
\begin{widetext}
\section{Superpositions of static trajectories}\label{appendixa}
For accessibility, we restate the static de Sitter metric here:
\begin{align}
    \D s^2 &= - f(r) \D t^2 + \frac{\D r^2}{f(r)} + r^2 \big( \D \theta^2 + \sin^2\theta \D \phi^2 \big) \qquad \text{where} \qquad f(r) = 1 - l^2r^2 . 
\end{align}
We now turn to the detector travelling in a superposition of trajectories in the static patch of de Sitter spacetime. We first consider a detector traveling on a superposition of worldlines defined by different $\theta$'s, with equal $l, r$. This corresponds to a spatial translation between the trajectories along an axis orthogonal to their motion. The trajectories are defined by the coordinates $\textbf{Z}^{ (A)} = (Z_0^{ (A)} , Z_i^{ (A)} )$ and $\textbf{Z}^{ (B)} = (Z_0^{ (B)}, Z_i^{ (B)})$,
\begin{align}
\begin{split}
    Z_0^{ (A)}&= \sqrt{l^{-2} - R_D^2} \sinh(lt) \\
    Z_1^{ (A)} &= \sqrt{l^{-2} - R_D^2}\cosh(lt) \\
    Z_2^{ (A)} &= R_D \cos\theta_A \vphantom{\sqrt{l^{-2} - R_D^2}} \\ 
    Z_3^{ (A)} &= R_D \sin\theta_A \cos\phi \vphantom{\sqrt{l^{-2} - r^2}} \\  
    Z_4^{ (A)} &= R_D \sin\theta_A \sin\phi \vphantom{\sqrt{l^{-2} - r^2}}.
\end{split}
\begin{split}
    Z_0^{ (B)} &= \sqrt{l^{-2} - R_D^2} \sinh(lt) \\
    Z_1^{ (B)} &= \sqrt{l^{-2} - R_D^2}\cosh(lt) \\
    Z_2^{ (B)} &= R_D \cos\theta_B \vphantom{\sqrt{l^{-2} - r^2}} \\ 
    Z_3^{ (B)} &= R_D \sin\theta_B \cos\phi \vphantom{\sqrt{l^{-2} - R_D^2}} \\ 
    Z_4^{ (B)} &= R_D \sin\theta_B \sin\phi \vphantom{\sqrt{l^{-2} - r^2}}.
\end{split}
\end{align}
The local Wightman functions are given by 
\begin{align}
    \mathcal{W}_{\mathcal{L}_S}^\text{loc.} (s) \equiv \mathcal{W}_\text{dS}(s) &= - \frac{\kappa^2}{16\pi^2} \frac{1}{\sinh^2(\kappa s/2 - i\varepsilon )},
\end{align}
while the nonlocal Wightman functions are given by 
\begin{align}\label{wightmantheta}
    \mathcal{W}^\text{int.}_{\mathcal{L}_S}(s) &= - \frac{\kappa^2}{16\pi^2} \frac{1}{\sinh^2(\kappa s/2 -i\varepsilon) - \kappa^2 R_D^2 \sin^2(\theta_S/2)}
\end{align}
where $\theta_S = \theta_A - \theta_B$. We have also defined $t = \tau /\sqrt{f(R_D)}$ and the proper time difference, $s = \tau - \tau'$. Equation (\ref{wightmantheta}) is time-translation invariant, reflecting the fact that the cosmological redshifts along each amplitude are equal. The simple form of the Wightman functions allows for analytical expressions to be obtained for the response function in the infinite interaction time limit. The response function integrals are
\begin{align}
    \mathcal{F}_{\mathcal{L}_S}^\text{loc.} (\Omega) &= - \frac{\kappa^2}{16\pi^2} \infint\D s \frac{e^{i\Omega s}}{\sinh^2(\kappa s/2 - i\varepsilon)} = \frac{\Omega}{2\pi} \frac{1}{e^{2\pi\Omega/\kappa} - 1} \\
    \mathcal{F}_{\mathcal{L}_S}^\text{int.}(\Omega) &= - \frac{\kappa^2}{16\pi^2} \infint\D s \:\frac{e^{i\Omega s }}{\sinh^2(\kappa s/2-i\varepsilon) - \kappa^2 R_D^2 \sin^2(\theta_S/2)} 
\end{align}
where the local term just yields the usual Planck distribution for a thermal bath radiating at the Gibbons-Hawking temperature $\kappa(2\pi)^{-1}$. For the interference terms, the poles in the denominator occur for $s = 2\pi n i\kappa^{-1} \pm 2\kappa^{-1} \sinh^{-1} ( \kappa R_D \sin\theta_S/2)$. The integral is thus\begin{align}
    \mathcal{F}_{\mathcal{L}_S}^\text{int.}(\Omega) &= \frac{\kappa^2}{16\pi^2}(2\pi i ) \sum_{n=-\infty }^{-1} \frac{e^{2\pi \Omega n /\kappa}\bigg\{ e^{-\frac{2i\Omega}{\kappa} \sinh^{-1} \big( R_D\kappa\sin\theta_S/2\big)  } - e^{\frac{2i\Omega}{\kappa} \sinh^{-1} \big( R_D\kappa \sin\theta_S/2 \big) } \bigg\} }{R_D\kappa^2 \sin(\theta_S/2) \sqrt{1 + R_D^2\kappa^2 \sin^2(\theta_S/2)}} .
    \intertext{Performing the summation in the lower half-plane yields}
   \mathcal{F}_{\mathcal{L}_S}^\text{int.}(\Omega) &= \frac{\Omega}{2\pi }\frac{1}{e^{2\pi \Omega/\kappa}-1} \frac{\sin \big( 2 \Omega\kappa^{-1} \text{arcsinh} \big( \kappa R_D \sin\theta_S/2 \big) \big)}{2\Omega R_D\sin(\theta_S/2) \sqrt{1 + \kappa^2 R_D^2\sin^2(\theta_S/2)}} = \frac{\Omega}{2\pi} \frac{1}{e^{2\pi \Omega/\kappa}-1} f(\Omega,\mathcal{L},\kappa)
\end{align} 
where
\begin{align} 
f(\Omega,\mathcal{L},\kappa) = \frac{\sin \big( 2 \Omega\kappa^{-1} \text{arcsinh} \big( \mathcal{L}_S\kappa/2 \big) \big)  }{\Omega \mathcal{L} \sqrt{1 + \big( \mathcal{L}_S\kappa/2 \big)^2}}
\end{align}
and we have defined $\mathcal{L}_S = 2R_D \sin \big( \theta_S/2 \big)$ as stated in the main text. Following this procedure, the response functions for the other stationary scenarios analysed can be straightforwardly derived. 

\section{Bipartite reduced density matrix in entanglement harvesting}\label{entanglementharvesting}
Here, we derive the bipartite reduced density matrix of two UdW detectors interacting with the quantum field in the protocol known as entanglement harvesting. The detectors travel on classical trajectories, but can extract nonlocal correlations through their interaction with the quantum field. The interaction Hamiltonian is given by 
\begin{align}
    \hat{H}_\text{int.} &= \sum_{D=A,B} \hat{\mathcal{H}}_{D}
\end{align}
where $D$ denotes the detector in question, and as before, 
\begin{align}
    \hat{\mathcal{H}}_{D} &= \lambda \eta_{D }(\tau) \sigma_D(\tau)  \hat{\Phi}(\mathsf{x}_{D} (\tau) )
\end{align}
describes the interaction of the $D$th detector with the quantum field along trajectory $\mathsf{x}_D(\tau)$. To second-order in perturbation theory, the time-evolution operator is
\begin{align}
    \hat{U} &= 1 - i\int\D \tau \:\big( \hat{\mathcal{H}}_{A}(\tau )  + \hat{\mathcal{H}}_{B}(\tau ) \big) - \iint_\mathcal{T} \D \tau \:\D \tau' \big( \hat{\mathcal{H}}_{A}(\tau )  + \hat{\mathcal{H}}_{B} ( \tau ) \big) \big( \hat{\mathcal{H}}_{A} ( \tau' ) + \hat{\mathcal{H}}_{B}(\tau' ) \big) .
\end{align}
The initial state of the system is simply
\begin{align}
    |\Psi\rangle_{CFD} &= |\psi\rangle_F \otimes  |g\rangle_A \otimes |g\rangle_B.
\end{align}
Evolving the state in time and tracing out the field degrees of freedom, it can be straightforwardly shown that the reduced density matrix of the two-detector system is given by 
\begin{align}\label{reduceddensity}
    \hat{\rho}_D &= \begin{pmatrix} 1 - \mathcal{P}_A -\mathcal{P}_B & 0 & 0 & \mathcal{M} \\ 0 & \mathcal{P}_B & \mathcal{L} & 0 \\ 0 & \mathcal{L}^\star  & \mathcal{P}_A & 0 \\ \mathcal{M}^\star & 0 & 0 & 0  \end{pmatrix}
\end{align}
where
\begin{align}
    \mathcal{P}_D &= \lambda^2 \iint\D \tau \:\D \tau' \big(\eta_{D}(\tau ) \eta_{D}(\tau' ) e^{-i\Omega(\tau - \tau' ) } \mathcal{W}\big(\mathsf{x}_{D}(\tau) ,\mathsf{x}_{D}(\tau' ) \big)  \\ 
    \mathcal{L} &= \lambda^2 \iint\D \tau \:\D\tau' \big( \eta_{B }(\tau)  \eta_{A}(\tau' ) e^{-i\Omega( \tau - \tau' ) } \mathcal{W} \big(\mathsf{x}_{A}(\tau), \mathsf{x}_{B}(\tau') \big) \\ 
    \mathcal{M} &= - \lambda^2 \iint_\mathcal{T} \D \tau \: \D \tau' e^{-i\Omega(\tau + \tau' ) } \big( \eta_{A }(\tau )  \eta_{B}(\tau' )  \mathcal{W} \big(\mathsf{x}_{A}(\tau ) , \mathsf{x}_{B}(\tau' ) \big) + \eta_{B} (\tau) \eta_{A}(\tau' ) \mathcal{W} \big(\mathsf{x}_{B } (\tau ) , \mathsf{x}_{A}(\tau' ) \big) \big) .
\end{align}
The advantage of entanglement harvesting in discerning the properties of the surrounding spacetime is that nonlocal correlations are encoded in the $\mathcal{L}$ and $\mathcal{M}$ terms of the bipartite reduced density matrix. However these nonlocal correlations, carried by the Wightman functions $\mathcal{W}\big( \mathsf{x}_A(\tau), \mathsf{x}_B(\tau') \big)$ and $\mathcal{W}\big(\mathsf{x}_B(\tau), \mathsf{x}_A(\tau') \big)$ are already accessible to a single UdW detector in a quantum superposition of paths, or residing in a spacetime in quantum superposition. The field operators are now evaluated along the trajectories $\mathsf{x}_A(\tau)$ and $\mathsf{x}_B(\tau)$ in superposition. 
\end{widetext}

\bibliography{References.bib}

\end{document}